\documentclass[prb,twocolumn,showkeys,amsmath,amsfonts]{revtex4-1}
\usepackage[latin2]{inputenc}
\usepackage{graphicx}
\usepackage{placeins}
\usepackage{color}
\usepackage{ulem}
\usepackage[colorlinks=true,citecolor=blue,linkcolor=magenta]{hyperref}
\usepackage{amsmath}
\def\be{\begin{equation}}
\def\ee{\end{equation}}
\def\bea{\begin{eqnarray}}          
\def\eea{\end{eqnarray}}
\def\bi{\begin{itemize}}
\def\ei{\end{itemize}}

\begin{document}

\title{    Variational tensor network renormalization in imaginary time:\\
           benchmark results in the Hubbard model at finite temperature 
}

\author{Piotr Czarnik} 
\affiliation{Instytut Fizyki Uniwersytetu Jagiello\'nskiego,
             ul. {\L}ojasiewicza 11, PL-30-348 Krak\'ow, Poland}
             
\author{Marek M. Rams} 
\affiliation{Instytut Fizyki Uniwersytetu Jagiello\'nskiego,
             ul. {\L}ojasiewicza 11, PL-30-348 Krak\'ow, Poland}             
 
\author{Jacek Dziarmaga} 
\affiliation{Instytut Fizyki Uniwersytetu Jagiello\'nskiego,
             ul. {\L}ojasiewicza 11, PL-30-348 Krak\'ow, Poland}

\date{\today}

\begin{abstract}
A Gibbs operator $e^{-\beta H}$ for a 2D lattice system with a Hamiltonian $H$ can be represented by a 3D tensor network, the third dimension being the imaginary time (inverse temperature) $\beta$. 
Coarse-graining the network along $\beta$ results in an accurate 2D projected entangled-pair operator (PEPO) with a finite bond dimension. 
The coarse-graining is performed by a tree tensor network of  isometries that are optimized variationally to maximize the accuracy of the PEPO. 
The algorithm is applied to the two-dimensional Hubbard model on an infinite square lattice.
Benchmark results are obtained that are consistent with the best cluster dynamical mean-field theory and numerically linked cluster expansion in the regime of parameters where they yield mutually consistent results.
\end{abstract}

\maketitle

%%%%%%%%%%%%%%%%%%%%%%%%%%%%%%%%%%%%%%%%%%%%%%%%%%%%%%%%%%%%%%%%%%%%%%%%%%
\section{Introduction}
%%%%%%%%%%%%%%%%%%%%%%%%%%%%%%%%%%%%%%%%%%%%%%%%%%%%%%%%%%%%%%%%%%%%%%%%%%

Quantum tensor networks prove to be an indispensable tool to study strongly correlated quantum systems. Their history begins with an introduction of the density matrix renormalization group (DMRG) \cite{White,Sch11} that was later shown to optimize the matrix product states (MPS) variational ansatz \cite{Sch11}. In recent years, MPS were generalized to 2D tensor product states, also known as a projected entangled pair states (PEPS) \cite{PEPS}, as well as supplemented with the multiscale entanglement renormalization ansatz (MERA) --- a refinement of the real-space renormalization group \cite{MERA}. Being variational, quantum tensor networks do not suffer form the notorious fermionic sign problem and thus they can be applied to strongly correlated fermions in 2D \cite{fermions1,fermions2}. The fermionic PEPS outperform the best variational quantum Monte Carlo (QMC) providing the most accurate results for the ground states of the $t-J$ \cite{ferVid,PEPStJ,CorbozQR,varCorboz} and Hubbard \cite{CorbozHubbard} models employed in the context of the high-${\cal T}_c$ superconductivity. The networks --- both MPS \cite{WhiteKagome,CincioVidal,starDMRG} and PEPS \cite{PepsRVB,PepsKagome,PepsJ1J2} --- also made major breakthroughs in the quest for topologically ordered ground states, where geometric frustration often prohibits traditional quantum Monte Carlo.

Thermal states of quantum Hamiltonians were explored with tensor networks to a lesser extent then the ground states. In 1D they can be represented by MPS prepared by accurate simulation of imaginary time evolution \cite{ancillas,WhiteT} of a system with ancillas. A similar approach can  be applied in 2D \cite{Czarniks,evolution,self} --- the PEPS manifold is a compact representation for Gibbs states \cite{Molnar} --- although quality of its results is inferior when compared with a method presented in this article. Alternatively, direct contraction of the 3D tensor network representing the partition function was proposed \cite{ChinaT}, but, due to local tensor update, they are expected to converge slowly with increasing refinement parameter. Any improvement towards the full update can accelerate the convergence significantly \cite{HOSRG}. These general efforts parallel similar progress in finite temperature variational quantum Monte Carlo, see e.g. Ref.~\onlinecite{japMC}.

Two of us introduced a variational algorithm to optimize a 2D projected entangled-pair operator (PEPO) representing the Gibbs operator $e^{-\beta H}$ for a 2D lattice Hamiltonian $H$ \cite{var,ourcompass} as an improvement of the imaginary time evolution approach. The variational method yields higher accuracy of the results and reduces significantly run time of the simulations, see Ref.~\onlinecite{var} and Appendix \ref{sec:imtev} for the comparison. 

Thanks to the area law for mutual information \cite{Molnar}, thermal states can be potentially represented by a PEPO with a finite bond dimension. Ref.~\onlinecite{Molnar} provides an upper bound for the bond dimension that is tight enough at high temperature.  While the above bound diverges at $T=0$, there is ample evidence that many ground states can be accurately represented by the PEPS ansatz with a finite bond dimension and a finite PEPO can be then obtained directly from PEPS in this limit, which allows us to anticipate the existence of sufficiently compact PEPO representation  for low temperatures as well.

The 2D PEPO is obtained by means of dimensional reduction from a 3D network that results from the Suzuki-Trotter decomposition of the infinitesimal time step $e^{-d\beta H}$, where $N=\beta/d\beta$ of such steps combine into the Gibbs operator. This decomposition is introducing a well controlled Trotter error which vanishes in the limit of $d\beta \to 0$. The 3D network is coarse-grained, or decimated, along the virtual (imaginary time) direction by a hierarchy of isometries. The isometries are optimized variationally -- employing full tensor environments -- in order to maximize the accuracy of the coarse-grained PEPO. Since the cost of the algorithm is only logarithmic in the number of time steps the limit of $d\beta \to 0$ can be easily approached for any finite $\beta$.

The benchmark application to the 2D quantum Ising model in a transverse field was presented in Ref.~\onlinecite{var}, while in Ref.~\onlinecite{ourcompass} the same algorithm was utilized to simulate the 2D quantum compass model \cite{Nus15} making the first step towards geometric frustration -- with the results consistent with the best QMC \cite{MCcompass2} ones. In this paper we present the first results for the strongly correlated fermionic Hubbard model. We limit ourselves to the regime where both numerically linked cluster expansion and cluster dynamical mean-field theory give mutually consistent results collected in Review~\onlinecite{Hubbardreview}, which are considered to be a benchmark for any new numerical approach.  Our results meet the standards of the test. 
 
The paper is organized as follows. 
We briefly define the Hubbard model in Sec.~\ref{sec:hubbard}, describe the fermionic algorithm for variational renormalization in Sec.~\ref{sec:algorithm}  and collect the numerical results in Sec.~\ref{sec:results}. We close with a discussion and conclusions in Sec.~\ref{sec:conclusion}.
In Appendix \ref{sec:CTM} we discuss the efficient modification of the corner transfer matrix algorithm which we employ in this work. Finally, in Appendix \ref{sec:imtev} we compare the accuracy of the algorithm with the earlier approach based on imaginary time evolution \cite{evolution} and in Appendix \ref{sec:spec} we elaborate on the convergence of the variational approach.

%%%%%%%%%%%%%%%%%%%%%%%%%%%%%%%%%%%%%%%%%%%%%%%%%%%%%%%%%%%%%%%%%%%%%%%%%%%%%%%%%%%%%%%%%%%
\section{Hubbard model}
\label{sec:hubbard}
%%%%%%%%%%%%%%%%%%%%%%%%%%%%%%%%%%%%%%%%%%%%%%%%%%%%%%%%%%%%%%%%%%%%%%%%%%%%%%%%%%%%%%%%%%%

The Hubbard model on an infinite square lattice is
\be
H = 
- \sum_{\langle i,j \rangle\sigma} 
  t\left( c_{i\sigma}^\dag c_{j\sigma} + c_{j\sigma}^\dag c_{i\sigma} \right) +  
  \sum_i U n_{i\uparrow} n_{i\downarrow}+
  \sum_i \mu\; n_i,
\label{H}
\ee
where $c_{i\sigma}^\dag$ and $c_{i\sigma}$, respectively,  creates and annihilates an electron with spin $\sigma=\uparrow,\downarrow$ on site $i$, $n_{i\sigma}=c^\dag_{i\sigma}c_{i\sigma}$ is the number operator, $n_i=n_{i\uparrow}+n_{i\downarrow}$,  on-site repulsion strength $U>0$ and the chemical potential is equal $\mu$. $\langle i,j \rangle$ denotes summation over nearest-neighbor (NN) pairs with hopping energy $t>0$. 
It is a candidate for a model describing high-$T_c$ superconductivity which is expected to appear close to $U/t=8$ and $n = 0.875$ at low enough temperatures. Here $n= \langle  n_i\rangle$ marks average electron density per site. 

For small $\beta$ the model was simulated using determinant quantum Monte Carlo (DQMC) \cite{Sca10} and dynamical cluster approximation (DCA) \cite{Sca07}. DCA is a variant of cluster dynamical mean-field theory (DMFT) using QMC impurity solver. Both approaches exhibit the sign problem which becomes more severe for growing both $\beta$ and the cluster size. It prevents from obtaining conclusive results for $\beta t > 4$ close to $U/t=8$ and $n = 0.875$  within these approaches \cite{Sca10,Sca07}.  Alternatively, the model can be simulated directly in the thermodynamic limit using numerically linked cluster expansion (NLCE)\cite{Kha11}. This method gives well controlled results for small $\beta t$ in the strongly correlated regime for the range of temperatures comparable with DQMC, but brakes down for larger  $\beta t$ \cite{Kha15}.   
In this article, as we are aiming to benchmark our approach, we work in the range of parameters when the above methods give mutually consistent results.

%%%%%%%%%%%%%%%%%%%%%%%%%%%%%%%%%%%%%%%%%%%%%%%%%%%%%%%%%%%%%%%%%%%%%%%%%% 
\section{ALGORITHM}
\label{sec:algorithm}
%%%%%%%%%%%%%%%%%%%%%%%%%%%%%%%%%%%%%%%%%%%%%%%%%%%%%%%%%%%%%%%%%%%%%%%%%% 

In this Section we describe fermionic version of the algorithm that was introduced and tested in the 2D quantum Ising model in Ref.~\onlinecite{var} and subsequently successfully applied to the quantum compass model in Ref.~\onlinecite{ourcompass}.

%%%%%%%%%%%%%%%%%%%%%%%%%%%%%%%%%%%%%%%%%%%%%%%%%%%%%%%%%%%%%%%%%%%%%%%%%% 

%%%%%%%%%%%%%%%%%%%%%%%%%%%%%%%%%%%%%%%%%%%%%%%%%%%%%%%%%%%%%%%%%%%%%%%%%% 
\subsection{Suzuki-Trotter decomposition}
\label{sec:ST} 
%%%%%%%%%%%%%%%%%%%%%%%%%%%%%%%%%%%%%%%%%%%%%%%%%%%%%%%%%%%%%%%%%%%%%%%%%%
We aim at representing the Gibbs operator,
\be 
U(\beta) = e^{-\beta H},
\ee
by a 2D tensor product operator shown in Figure \ref{fig:Tm}c,  also known as projected entangled-pair 
operator (PEPO). To this end, we first split the evolution into $N$ small time steps
\be 
U(\beta)=\left[U(d\beta)\right]^N,
\label{UN}
\ee
where
\be 
N=\frac{\beta}{d\beta}
\ee
is the number of steps. Since the cost of our algorithm scales with $\log_2N$, 
the step $d\beta$ can be made sufficiently small at little expense.

Since the Hamiltonian (\ref{H}) is a sum of non-commuting terms, 
the gate $U(d\beta)=e^{-d\beta H}$ has to be approximated using the 
Suzuki-Trotter decomposition by a product of  
non-commuting Trotter gates. The on-site terms in (\ref{H}) act with a gate,
\be 
U_{U}(d\beta)=\prod_i 
e^{-d\beta\left( U n_{i\uparrow}n_{i\downarrow}+\mu n_i\right)},
\label{UU}
\ee 
being a product of one-site operators.
In order to implement the hopping
we introduce two sublattices, $A$ and $B$, which form a chessboard and define hopping gates:
\bea
U^\sigma_{AB}(d\beta) &=&
\prod_{\langle i_A,i_B\rangle}e^{td\beta~ c^\dag_{i_A\sigma}c_{i_B\sigma} },\\
U^\sigma_{BA}(d\beta) &=&
\prod_{\langle i_A,i_B\rangle}e^{td\beta~ c^\dag_{i_B\sigma}c_{i_A\sigma} }.
\eea
Here $\langle i_A,i_B\rangle$ marks a pair of NN sites belonging to different sublattices and
the gates are products of NN two-site commuting operators.
In the second-order Suzuki-Trotter decomposition an infinitesimal imaginary time step is approximated by
\be
U(d\beta) \approx
U_U\left(d\beta/2\right)
U^\uparrow_t\left(d\beta\right)
U^\downarrow_t\left(d\beta\right)
U_U\left(d\beta/2\right),
\label{Udbeta}
\ee
where the hopping gates are combined into two mutually commuting gates
\be 
U^\sigma_t(d\beta) =
U^\sigma_{AB}\left(d\beta/2\right)
U^\sigma_{BA}\left(d\beta\right)
U^\sigma_{AB}\left(d\beta/2\right).
\label{Ut}
\ee
In order to rearrange $U^\sigma_{AB}(d\beta)$ in the form of a tensor network, 
at every NN bond we rewrite exactly:
\be 
e^{t d\beta~ c^\dag_{i_A\sigma}c_{i_B\sigma} } =
1+t d \beta ~ c^\dag_{i_A\sigma}c_{i_B\sigma} =
\sum_{\nu=0,1} C^{\nu\dag}_{i_A\sigma d\beta} C^\nu_{i_B\sigma d\beta} .
\ee
Above, $C^{\nu(\dag)}_{i\sigma d\beta}=\left(c^{(\dag)}_{i\sigma}\sqrt{td\beta}\right)^\nu$ is acting on site $i$ and the bond index $\nu$ contracts two operators acting on the NN sites $i_A$ and  $i_B$. In the Fock basis at the site $i$ this operator becomes a rank-3 tensor $C^{(\dag)}_{\sigma d\beta}$ with elements:
\be
C^{(\dag)\nu mn}_{i\sigma d\beta}=\langle m| C^{\nu(\dag)}_{i\sigma d\beta} |n\rangle.
\label{Ctensor}
\ee
We show its graphical representation in Fig.~\ref{fig:SWAP}a.

\subsection{Fermionic tensor networks}

In order to represent fermionic states by tensor networks we use formalism of Ref.~\onlinecite{ferVid}. We introduce it shortly here, while referring to Ref.~\onlinecite{ferVid} for a  systematic derivation.

 We assume that  each value of each  tensor index has even or odd parity, e.g., for $C^{\nu mn}_{i\sigma d\beta}$,    
 the indices $m,n$ are even (odd) when they label a Fock state with even (odd) number of fermions, and even (odd) bond index $\nu=0(1)$ corresponds to a transfer of an even (odd) number of fermions between the NN sites. 
It is convenient to denote the parity of an index $i$  by 
\begin{equation}
p(i) = \begin{cases}
1 & \textrm{for }\,i\,\textrm{ even}, \\
-1 &  \textrm{for }\,i\,\textrm{ odd}. \\
\end{cases}
\end{equation}
Then, the parity of a set of indices is defined as product of their parities, 
\begin{equation}
p(\{i_1,i_2\}) = p(i_1)p(i_2).
\end{equation}
%%%%%%%%%%%%%%%%%%%%%%%%%%%%%%%%%%%%%%%%%%%%%%%%%%%%%%%%%%%%%%%%%%%%%%%%%%%%
\begin{figure}[t!]
\vspace{-0cm}
\includegraphics[width=0.99\columnwidth,clip=true]{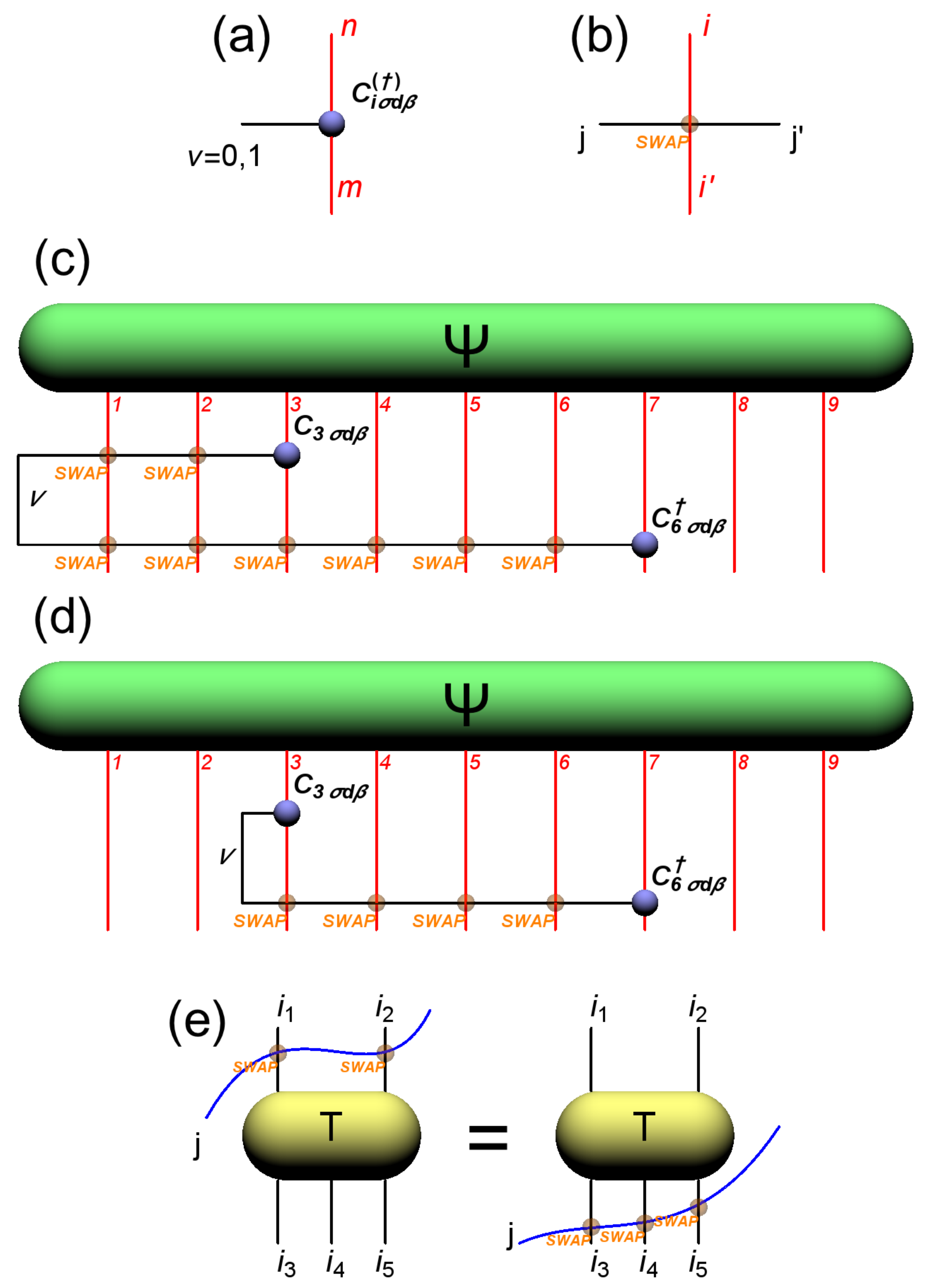}
\vspace{-0cm}
\caption{ 
In (a), 
graphical representation of fermionic tensor $C^{(\dag)}_{i\sigma d\beta}$ with elements
$
C^{(\dag)\nu mn}_{i\sigma d\beta}=\langle m| C^{\nu(\dag)}_{i\sigma d\beta} |n\rangle
$ 
for site $i$ and spin $\sigma$.
The black line represents the bond index $\nu$ connecting this operator with a NN site along the bond.
The red lines represent indices $m$ and $n$ numbering Fock states at site $i$. 
In (b), graphical representation of the swap gate  $s_{i,j,i',j'} = \delta_{i,i'} \delta_{j,j'} S(i,j)$ where $S(i,j) = -1 $ if both values of indices $i$ and $j$ are odd and $S(i,j) = 1 $ otherwise.
In (c),
the hopping gate $\sum_{\nu}C^{\nu\dag}_{6\sigma d\beta}C^{\nu}_{3\sigma d\beta}$ is applied to a quantum state $|\Psi\rangle$. At every crossing of the  bond index $\nu=0,1$ (black line) with a  fermionic index (red lines), which numbers Fock states, there is a fermionic swap gate. The gate produces  factor $-1$ when both values of the crossing indices are odd and factor $1$  otherwise. Here, each of the red indices is odd when it labels a Fock state with an odd number of fermions. The bond index is odd when it labels a transfer of an odd number of fermions between sites ($\nu=1$ here). 
In (d), 
the pairs of swap gates at sites $1$ and $2$ cancel out  so the bond line can be shortened. 
In (e), we show
an example of freedom in choosing  tensor network representations of fermionic systems. The index $j$ (the blue line) can be dragged over a parity  preserving tensor $T$ without changing factors produced by the swap gates.  }
\label{fig:SWAP}
\end{figure}
%%%%%%%%%%%%%%%%%%%%%%%%%%%%%%%%%%%%%%%%%%%%%%%%%%%%%%%%%%%%%%%%%%%%%%%%%%%%
In the following we work only with the parity preserving tensors, i.e., 
\begin{equation}
T_{i_1,i_2,\dots,i_r} = 0 \quad \textrm{if} \quad p(\{i_1,i_2,\dots,i_r\}) = -1.
\end{equation}

%%%%%%%%%%%%%%%%%%%%%%%%%%%%%%%%%%%%%%%%%%%%%%%%%%%%%%%%%%%%%%%%%%%%%%%%%% 
%\subsection{Fermionic swap gates}
%\label{sec:SWAP} 
%%%%%%%%%%%%%%%%%%%%%%%%%%%%%%%%%%%%%%%%%%%%%%%%%%%%%%%%%%%%%%%%%%%%%%%%%%

A new feature of fermionic tensor networks are fermionic swap gates which implement fermionic anticommutators. We illustrate the general idea with an example, while a more systematic introduction can be found in Ref.~\onlinecite{ferVid}. Fig.~\ref{fig:SWAP}c shows the hopping term 
$
\sum_{\nu=0,1} C_{6\sigma d\beta}^{\dag\nu} C_{3\sigma d \beta}^{\nu}
$
from site $3$ to site $6$ applied to a state $|\Psi\rangle$, where the rank-$9$ tensor $\Psi$ represents  its  amplitude in the Fock basis:
\be 
|\Psi\rangle=
\sum_{n_1,...,n_9=0,1}
\Psi_{n_1,...,n_9}
\left(c_{1\sigma}^\dag\right)^{n_1}...
\left(c_{9\sigma}^\dag\right)^{n_{9}}
|0\rangle.
\ee 
For odd $\nu=1$ we have $C_{3\sigma d\beta}^{\nu}\sim c_{3\sigma}$ and the intermediate state $c_{3\sigma}|\Psi\rangle$ is
\bea 
\sum_{n_1,...,n_9=0,1}
&&
\Psi_{n_1,...,n_9} ~
(-1)^{n_1}(-1)^{n_2} ~
\delta_{1,n_3}    \nonumber\\
&&
\left(c_1^\dag\right)^{n_1}
\left(c_2^\dag\right)^{n_2}
\left(c_3^\dag\right)^{0}...
\left(c_9^\dag\right)^{n_9}
|0\rangle.
\label{ex1}
\eea
Thanks to fermionic anticommutators, the amplitude acquires a  factor $(-1)^{n_1}(-1)^{n_2}$ which in the tensor network formalism can be taken into account by introduction of swap gates.   A swap gate is a rank 4 tensor $s_{i,j,i',j'}$,
\begin{equation}
s_{i,j,i',j'} = \delta_{i,i'} \delta_{j,j'} S(i,j),
\end{equation}
where
\begin{equation}
S(i,j) = 
\begin{cases}
-1 & \text{for } p(i) = p(j) = -1, \\
1 & \text{otherwise.} \\
\end{cases}
\end{equation}  
Its graphical representation is shown in Fig.~\ref{fig:SWAP}b.  

The factor $(-1)^{n_1}(-1)^{n_2}$ from Eq.~\eqref{ex1}  (see Fig.~\ref{fig:SWAP}c)  is represented by the two swap gates appearing at the crossing of the bond-index $\nu$  (black lines in the figure)  coming from the tensor $C_{3\sigma d\beta}$ and the  fermionic  indices at sites $1$ and $2$ (red lines).  For 
$\nu=1$ the two gates in Fig.~\ref{fig:SWAP}c produce the required factor.
The action of the hopping term is completed by applying $C_{6\sigma d\beta}^{\dag\nu}$ at site $6$. Its application produces, for $\nu=1$, another factor $(-1)^{n_1}(-1)^{n_2}\dots (-1)^{n_6}$.  This factor is represented in Fig.~\ref{fig:SWAP}c by swap gates at the crossings of  the (black) bond index $\nu$  and (red) fermionic  indices $1,2,\dots,6$.

To summarize, for $\nu=1$ we obtained in total two factors $(-1)^{n_1}$ and two factors $(-1)^{n_2}$ which, respectively, cancel out. This  corresponds to  cancellation of two swap gates  between the same pairs of indices. Therefore, we can remove swap gates at the crossings of the fermionic indices $1$ and $2$ with the  bond index $\nu$, and obtain the final tensor network representation of the hopping term  in Fig.~ \ref{fig:SWAP}{d}.

\begin{figure}[t!]
\vspace{-0cm}
\includegraphics[width=0.99\columnwidth,clip=true]{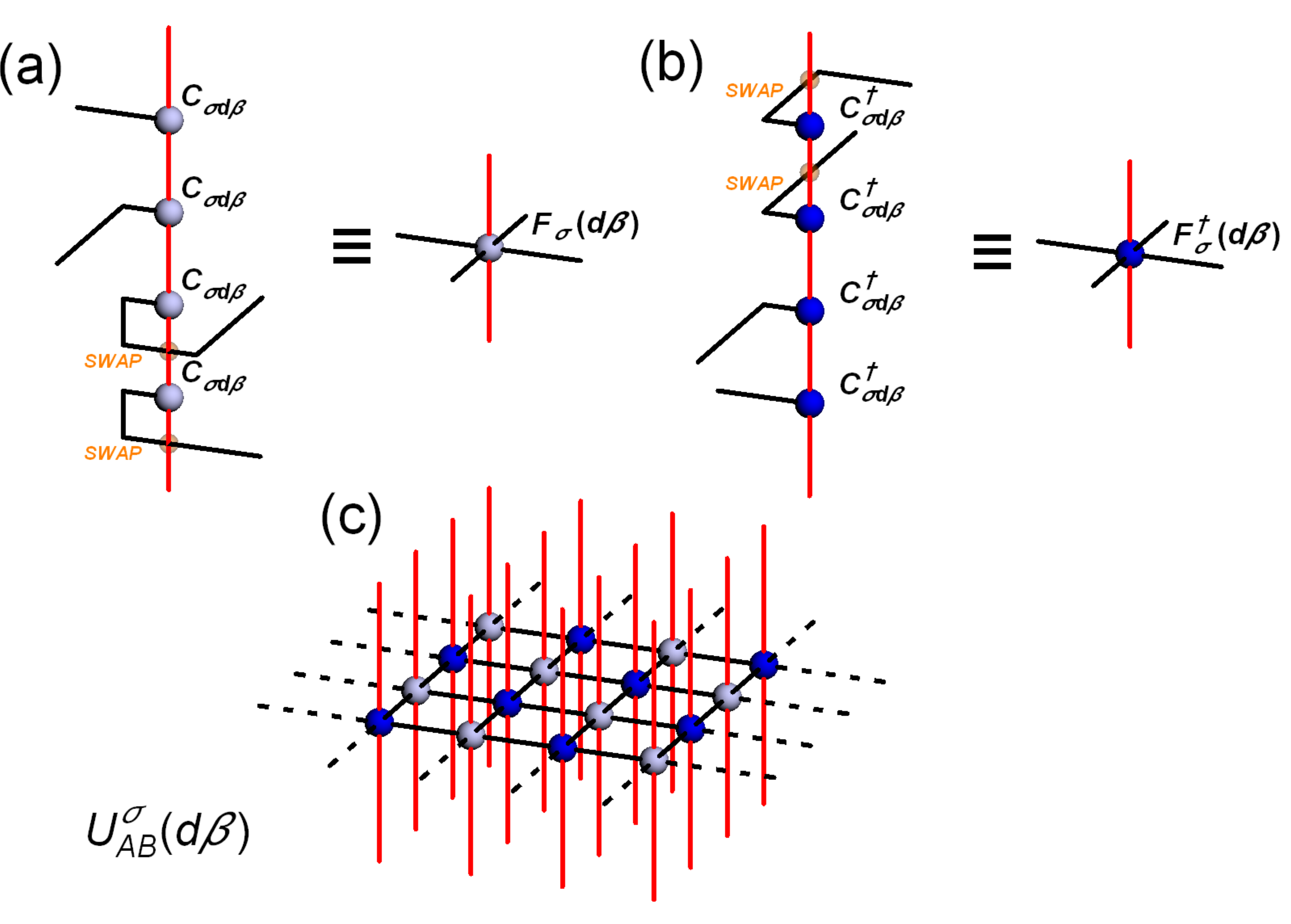}
\vspace{-0cm}
\caption{ 
In (a), 
four operators $C^\nu_{i\sigma d\beta}$ are applied at site $i$. Each operator has one bond index connecting it with an operator $C^{\nu\dag}_{j\sigma d\beta}$ at a NN site $j$. 
The twisted bond lines respect the Fock ordering of sites assumed on the 2D square lattice, 
compare Fig.~\ref{fig:SWAP}c and d in case of 1D.
The whole diagram can be represented by a compact rank-$6$ tensor $F_{\sigma}(d\beta)$, where
the twisted bonds are hidden in its internal structure.
In (b)
four operators $C^{\nu\dag}_{i\sigma d\beta}$ are represented by a rank-6 tensor 
$F^\dag_{\sigma}(d\beta)$. 
In (c)
the gate $U^\sigma_{AB}(d\beta)$ in Eq.~\eqref{UAB} is a chessboard constructed from $F_{\sigma}(d\beta)$ and $F^\dag_{\sigma}(d\beta)$, where each line connecting two NN tensors at sites $i_A$ and $i_B$ represents contraction of their common bond index $\nu_{i_Ai_B}$ and each line crossing implies a swap factor. 
The internal structure of the tensors $F_{\sigma}(d\beta)$ and $F^\dag_{\sigma}(d\beta)$ keeps track of the assumed Fock ordering of lattice sites. 
}
\label{fig:CC}
\end{figure}

In the following  any crossing of two lines implies a swap gate between the corresponding indices.  For the sake of clarity, starting from Figure \ref{fig:CC}c on we do not show the swap gates explicitly.

Finally, Fig.~\ref{fig:SWAP}e illustrates the advantage of using parity preserving tensors. In Fig.~\ref{fig:SWAP}e a parity preserving tensor $T_{i_1,i_2,i_3,i_4,i_5}$ is swapped with an index $j$ (the blue line) in two equivalent ways. The left and right  diagrams denote respectively  $\sum_{i_1',i_2',j'} T_{i'_1,i'_2,i_3,i_4,i_5}s_{i_1,j,i_1',j'}s_{i_2,j',i_2',j''}$ and $\sum_{i_3',i_4',i_5',j',j'',} T_{i_1,i_2,i'_3,i'_4,i'_5}s_{i_3,j,i'_3,j'}s_{i_4,j',i_4',j''}s_{i_5,j'',i'_5,j'''}$.  Because $T_{i_1,i_2,i_3,i_4,i_5}$ is a parity preserving tensor then, by definition, it is only nonzero for $p({i_1,i_2}) = p({i_3,i_4,i_5})$ which  in turns implies that for each given value of $j$ index a factor introduced by the swap gates between $j$ index and $i_1$,$i_2$ indices is the same as a factor introduced by the swap gates between  $j$ index and $i_3$,$i_4$,$i_5$ indices.  Therefore we can drag the blue line over the tensor. In the following we frequently drag indices over tensors to optimize cost of tensor contractions.

%%%%%%%%%%%%%%%%%%%%%%%%%%%%%%%%%%%%%%%%%%%%%%%%%%%%%%%%%%%%%%%%%%%%%%%%%%%%
%%%%%%%%%%%%%%%%%%%%%%%%%%%%%%%%%%%%%%%%%%%%%%%%%%%%%%%%%%%%%%%%%%%%%%%%%%%%

%%%%%%%%%%%%%%%%%%%%%%%%%%%%%%%%%%%%%%%%%%%%%%%%%%%%%%%%%%%%%%%%%%%%%%%%%% 
\subsection{Hopping gates}
\label{sec:hop} 
%%%%%%%%%%%%%%%%%%%%%%%%%%%%%%%%%%%%%%%%%%%%%%%%%%%%%%%%%%%%%%%%%%%%%%%%%%

Now we can proceed with rewriting the hopping gate as a contraction of tensors via their bond indices:
\bea
U^\sigma_{AB}(d\beta)&=&
\prod_{\langle i_A,i_B\rangle}e^{d\beta~ c^\dag_{i_A\sigma}c_{i_B\sigma} }=
\nonumber\\
&&
\sum_{\{\nu\}}
\prod_{\langle i_A,i_B \rangle}
C_{i_A\sigma d\beta}^{\dag \nu_{i_Ai_B}}
C_{i_B\sigma d\beta}^{\nu_{i_Ai_B}}.
\label{UAB}
\eea
Here $\nu_{i_Ai_B}$ is a bond index along NN bond $\langle i_A,i_B\rangle$ and $\{\nu\}$ is a set of all such bond indices. 
In Fig.~\ref{fig:CC}a, four operators $C^\nu_{i_B\sigma d\beta}$ are combined into a single rank-$6$ tensor $F_{\sigma}(d\beta)$. In Fig.~\ref{fig:CC}b, four operators $C^{\nu\dag}_{i_A\sigma d\beta}$ are combined into $F_{\sigma}^\dag(d\beta)$. In Fig.~\ref{fig:CC}c, a chessboard layer of $F_{\sigma}(d\beta)$ and $F_{\sigma}^\dag(d\beta)$ is shown representing the gate $U^\sigma_{AB}(d\beta)$. The other gate $U^\sigma_{BA}(d\beta)$ is a similar layer but with $F_{\sigma}(d\beta)$ and $F_{\sigma}^\dag(d\beta)$ interchanged. 

%%%%%%%%%%%%%%%%%%%%%%%%%%%%%%%%%%%%%%%%%%%%%%%%%%%%%%%%%%%%%%%%%%%%%%%%%%%%
\begin{figure}[t!]
\vspace{-0cm}
\includegraphics[width=0.99\columnwidth,clip=true]{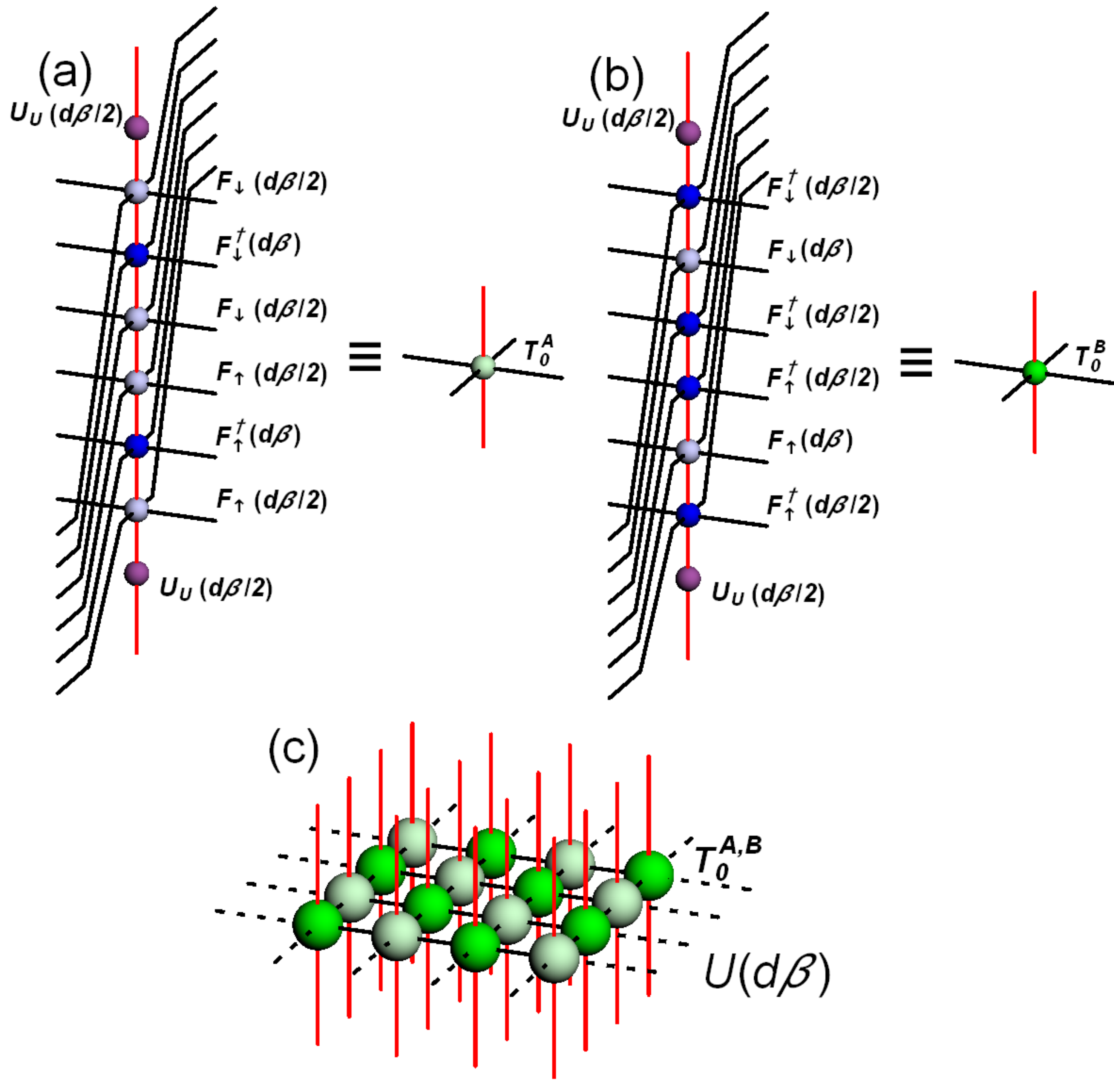}
\vspace{-0cm}
\caption{ 
In (a),
$F$ and $F^\dag$ on sublattice $A$ are combined into a single elementary Trotter tensor $T^A_0$ 
according to Eqs. (\ref{Udbeta},\ref{Ut}). Each bond index of new $T^A_0$ has dimension $2^6$.
The internal structure of tensors $F$ and $F^\dag$ keeps track of the original Fock ordering of 
lattice sites.
In (b),
similar operation is performed for sublattice $B$.
In (c),
a layer of the elementary Trotter tensors forms the infinitesimal gate $U(d\beta)$ and
$N$ such layers put on top of each other represent the Gibbs operator 
$U(\beta)$ in Eq.~\eqref{UN}.   
}
\label{fig:T0}
\end{figure}
%%%%%%%%%%%%%%%%%%%%%%%%%%%%%%%%%%%%%%%%%%%%%%%%%%%%%%%%%%%%%%%%%%%%%%%%%%%%

According to Eq.~\eqref{Ut}, a sandwich of three such chessboard layers on top of each other is the hopping gate $U_t^\sigma$. Following from Eq.~\eqref{Udbeta}, two such sandwiches on top of each other, dressed at both ends with the on-site gates $U_U$, combine into the infinitesimal gate $U(d\beta)$. In practice,
it is not possible to contract such 3D layered networks exactly and, therefore, we proceed by contracting them along their Fock indices first. This is presented in Figs.~\ref{fig:T0}a and b, where the operators corresponding to sublattices $A$ and $B$ are combined into elementary Trotter tensors $T_0^A$ and 
$T_0^B$, respectively. To that end, the $6$ bond indices along each direction on the lattice are fused together into a single bond index of dimension $2^6$. A layer of such tensors, shown in Fig.~\ref{fig:T0}c, represents the infinitesimal gate $U(d\beta)$ in Eq.~\eqref{UAB}, up to the Trotter error controlled by $d \beta$. Finally,  the evolution operator $U(\beta)$  in Eq.~\eqref{UN} is obtained by placing  $N$ such layers on top of each other.

%%%%%%%%%%%%%%%%%%%%%%%%%%%%%%%%%%%%%%%%%%%%%%%%%%%%%%%%%%%%%%%%%%%%%%%%%%%%
\begin{figure}[t!]
\vspace{-0cm}
\includegraphics[width=0.6\columnwidth,clip=true]{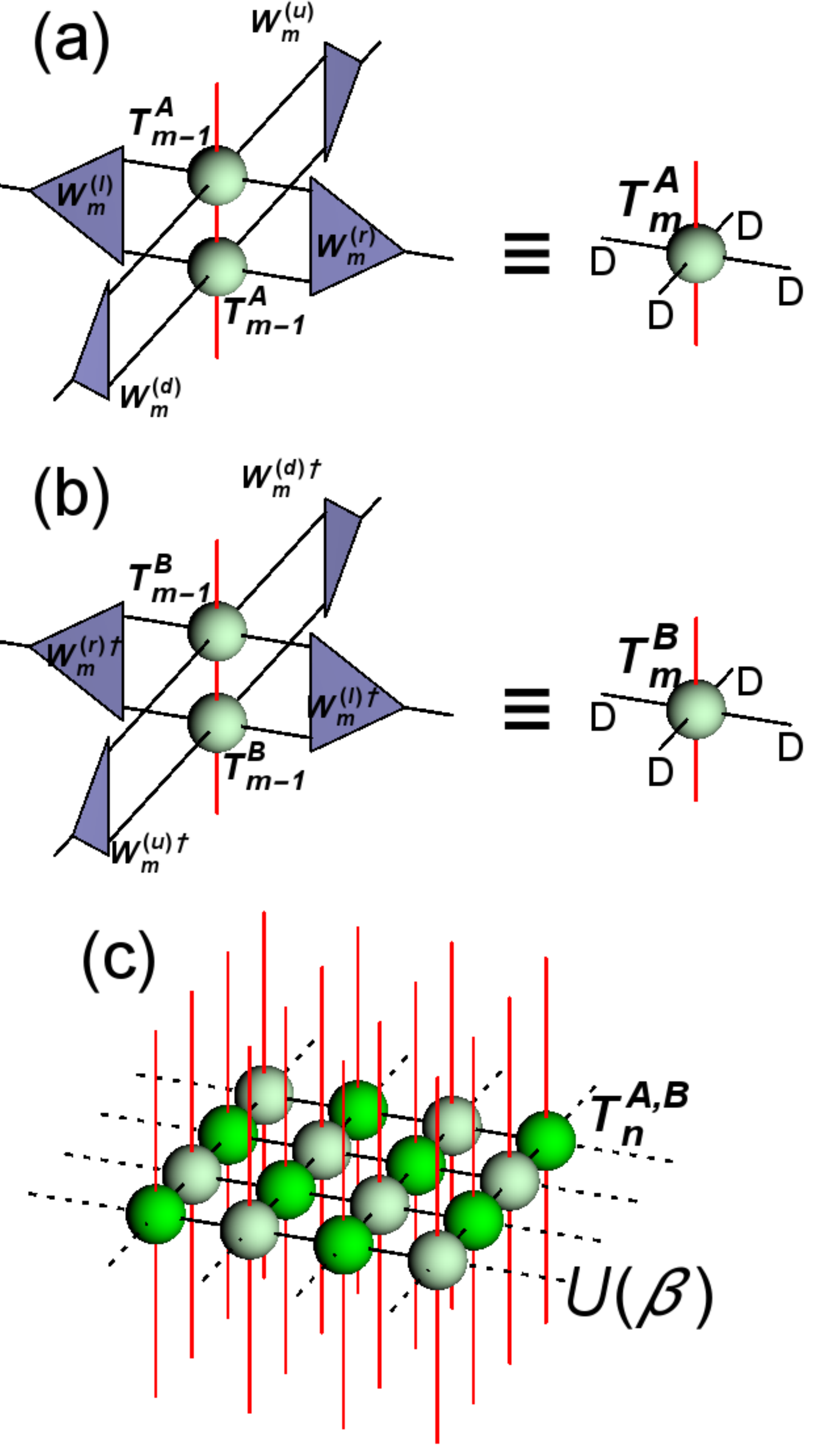}
\vspace{-0cm}
\caption{
In (a),
a coarse-graining step along virtual (imaginary time) direction, 
where two Trotter tensors $T^A_{m-1}$ are combined and 
renormalized into a single tensor $T^A_m$ by four isometries $W_m^{(u)},W_m^{(r)},W_m^{(d)},W_m^{(l)}$.  
Similarly, in (b),
isometries $W_m^{(u)\dag},W_m^{(r)\dag},W_m^{(d)\dag},W_m^{(l)\dag}$ renormalize corresponding indices of a combination $T^B_{m-1}\times T^B_{m-1}\to T^{B}_m$. 
Finally, PEPO tensors $T_n^{A}$ and $T_n^{B}$ result from $n$ such coarse-graining transformations.
In (c), 
a chessboard layer of contracted $T_n^{A,B}$ forms the PEPO ansatz for the Gibbs operator $U(\beta)$. 
}
\label{fig:Tm}
\end{figure}
%%%%%%%%%%%%%%%%%%%%%%%%%%%%%%%%%%%%%%%%%%%%%%%%%%%%%%%%%%%%%%%%%%%%%%%%%%%%%

%%%%%%%%%%%%%%%%%%%%%%%%%%%%%%%%%%%%%%%%%%%%%%%%%%%%%%%%%
\begin{figure}[!]
\vspace{-0cm}
\includegraphics[width=0.99\columnwidth,clip=true]{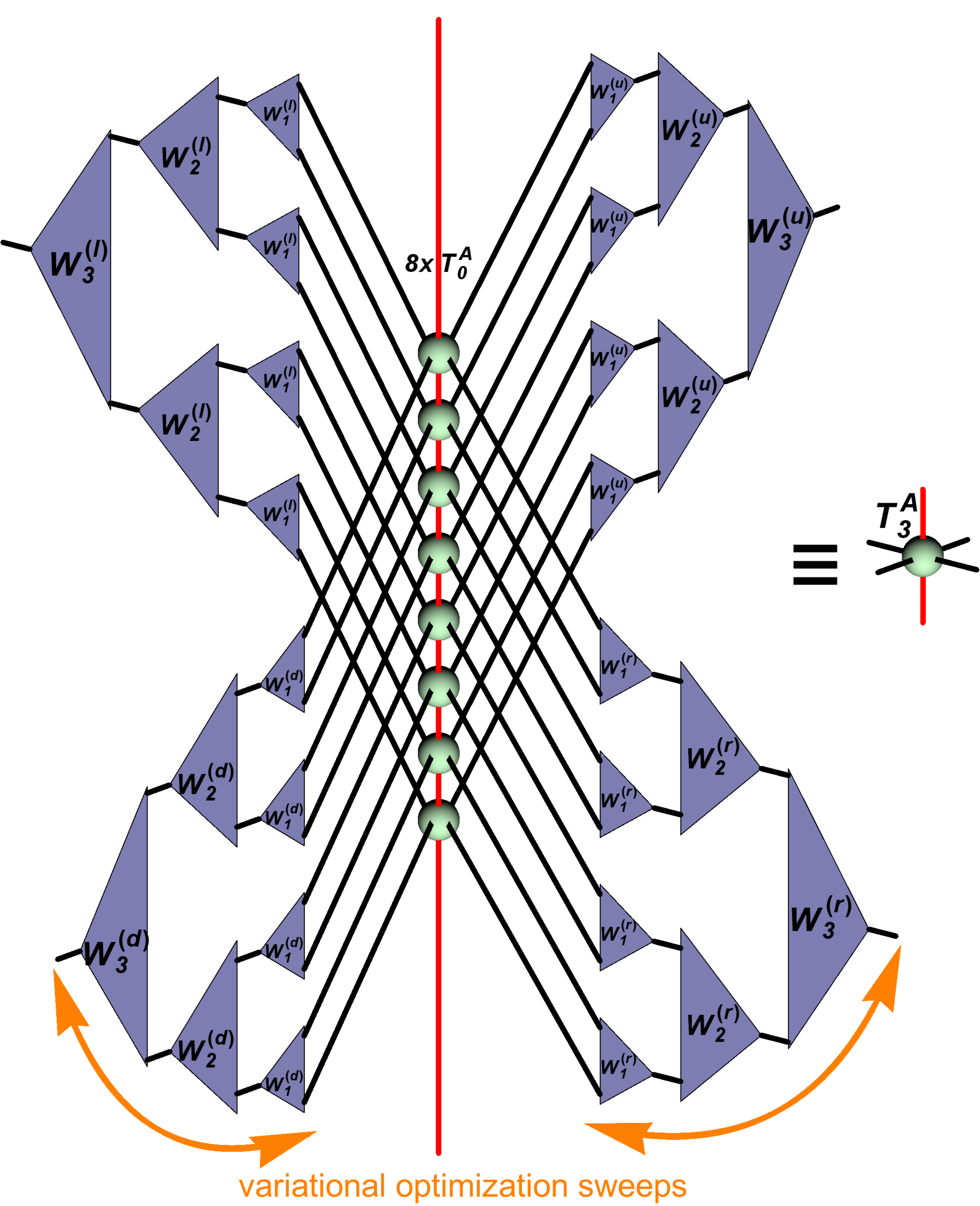}
\vspace{-0cm}
\caption{
Three coarse-graining transformations in Fig.~\ref{fig:Tm}c result in a Trotter tensor $T^A_3$.
After  dragging the lines  coming from the eight elementary $T^A_0$ over the isometries,
like in Fig.~\ref{fig:SWAP}e,
the isometries acting along a given bond combine into a tree tensor network (TTN). 
They are optimized by repeated up- and down-sweeps.  
}
\label{fig:TTN}
\end{figure}
%%%%%%%%%%%%%%%%%%%%%%%%%%%%%%%%%%%%%%%%%%%%%%%%%%%%%%%%%%%%%%%%%%%%%%%%%%%% 

%%%%%%%%%%%%%%%%%%%%%%%%%%%%%%%%%%%%%%%%%%%%%%%%%%%%%%%%%%%%%%%%%%%%%%%%%% 
\subsection{Coarse graining and renormalization in imaginary time}
\label{sec:CG} 
%%%%%%%%%%%%%%%%%%%%%%%%%%%%%%%%%%%%%%%%%%%%%%%%%%%%%%%%%%%%%%%%%%%%%%%%%%

Equation \eqref{UN} suggests to combine $N$ elementary tensors $T^{A}_0$ or $T^{B}_0$ into a single PEPO tensor in a similar way as in Figs.~\ref{fig:T0}a or b, where six tensors were combined into a single Trotter tensor $T^A_0$ or $T^B_0$. Unfortunately, we cannot proceed along these lines anymore as this would require fusing $N$ bond indices along each lattice direction, each of dimension $2^6$, into a single bond index  of dimension $2^{6N}$.

In order to avoid this exponential blow-up in $N$, we prefer to proceed step-by-step, each time combining just two tensors into one and renormalizing the bond indices:
$T_0^{A,B}\times T_0^{A,B}\to T_1^{A,B}$,...,$T_{n-1}^{A,B}\times T_{n-1}^{A,B}\to T_n^{A,B}$,
where 
\be 
n=\log_2N=\log_2 \frac{\beta}{d\beta}
\ee
is the total number of transformations. It is scaling logarithmically in the total number of Suzuki-Trotter steps $N$ or, equivalently, in $d\beta$. 

After each fusion of two tensors, $T_{m-1}^{A,B}\times T_{m-1}^{A,B}\to T_m^{A,B}$, 
the two bond indices along each lattice direction are renormalized down to $D$ using four isometries: 
$W_m^{(u)},W_m^{(r)},W_m^{(b)},W_m^{(l)}$, as shown in Figs.~\ref{fig:Tm}a and b. This renormalization keeps the bond dimension fixed at a tractable $D$.
Figure \ref{fig:TTN}, for instance, shows the net outcome after $3$ such coarse-graining 
transformations. Along each bond there are $3$ layers of isometries:
$W_1,W_2,W_3$ that combine into structure of a tree tensor network (TTN) \cite{TTN}.

%%%%%%%%%%%%%%%%%%%%%%%%%%%%%%%%%%%%%%%%%%%%%%%%%%%%%%%%%%%%%%%%%%%%%%%%%%%
\begin{figure}[t!]
\vspace{-0cm}
\includegraphics[width=0.7\columnwidth,clip=true]{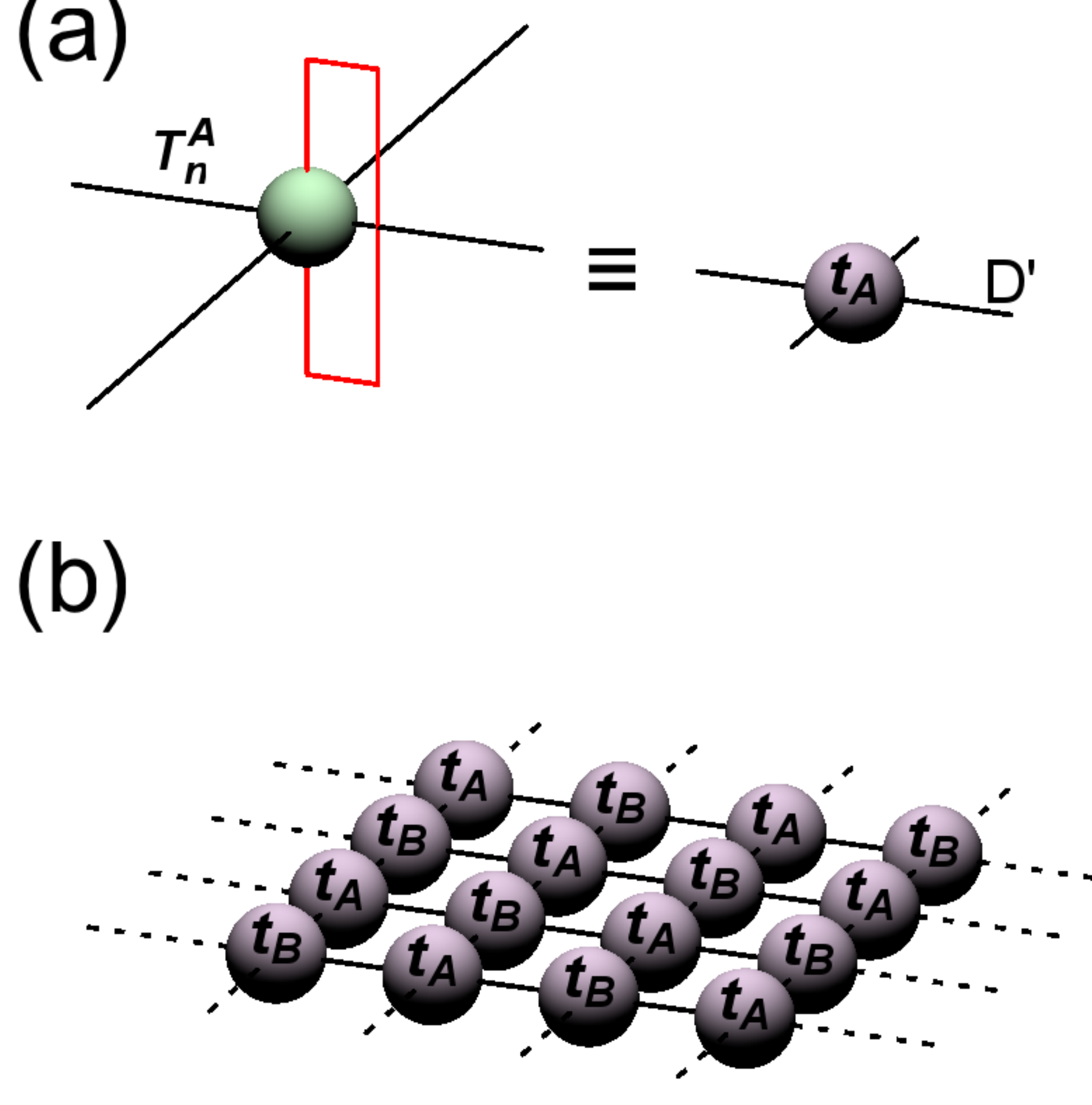}
\vspace{-0cm}
\caption{
In (a),
tracing out the (red) fermionic indices of tensor $T_n^A$ forms a transfer tensor $t_A$.
In (b), 
chessboard layer of contracted transfer tensors $t_A$ and $t_B$ represents the partition 
function $Z={\rm Tr}\,U(\beta)={\rm Tr}\,e^{-\beta H}$.
}
\label{fig:Tn}
\end{figure}
%%%%%%%%%%%%%%%%%%%%%%%%%%%%%%%%%%%%%%%%%%%%%%%%%%%%%%%%%%%%%%%%%%%%%%%%%%

After $n$ coarse-graining steps we obtain tensors $T_n^{A,B}$. 
Their chessboard layer shown in Fig.~\ref{fig:Tm}c is the PEPO ansatz for the gate $U(\beta)$. 
In the final $n$-th renormalization we  insert   a trivial isometry (the  identity) increasing the bond dimension  to 
\be
D'=D^2
\ee
to minimize effects of the final truncation.
  
The partition function $Z={\rm Tr}\,e^{-\beta H}$ is a trace of the PEPO.
It can be represented by a chessboard in Fig.~\ref{fig:Tn}b formed by transfer tensors $t_A$ and $t_B$. 
The transfer tensors are obtained by tracing out fermionic indices of corresponding
PEPO tensors $T_n^{A,B}$, see Fig.~\ref{fig:Tn}a.
These rank-4 transfer tensors have only bond indices with the bond dimension $D'$ .

%%%%%%%%%%%%%%%%%%%%%%%%%%%%%%%%%%%%%%%%%%%%%%%%%%%%%%%%%%%%%%%%%%%%%%%%%%%
\begin{figure}[t!]
\vspace{-0cm}
\includegraphics[width=0.7\columnwidth,clip=true]{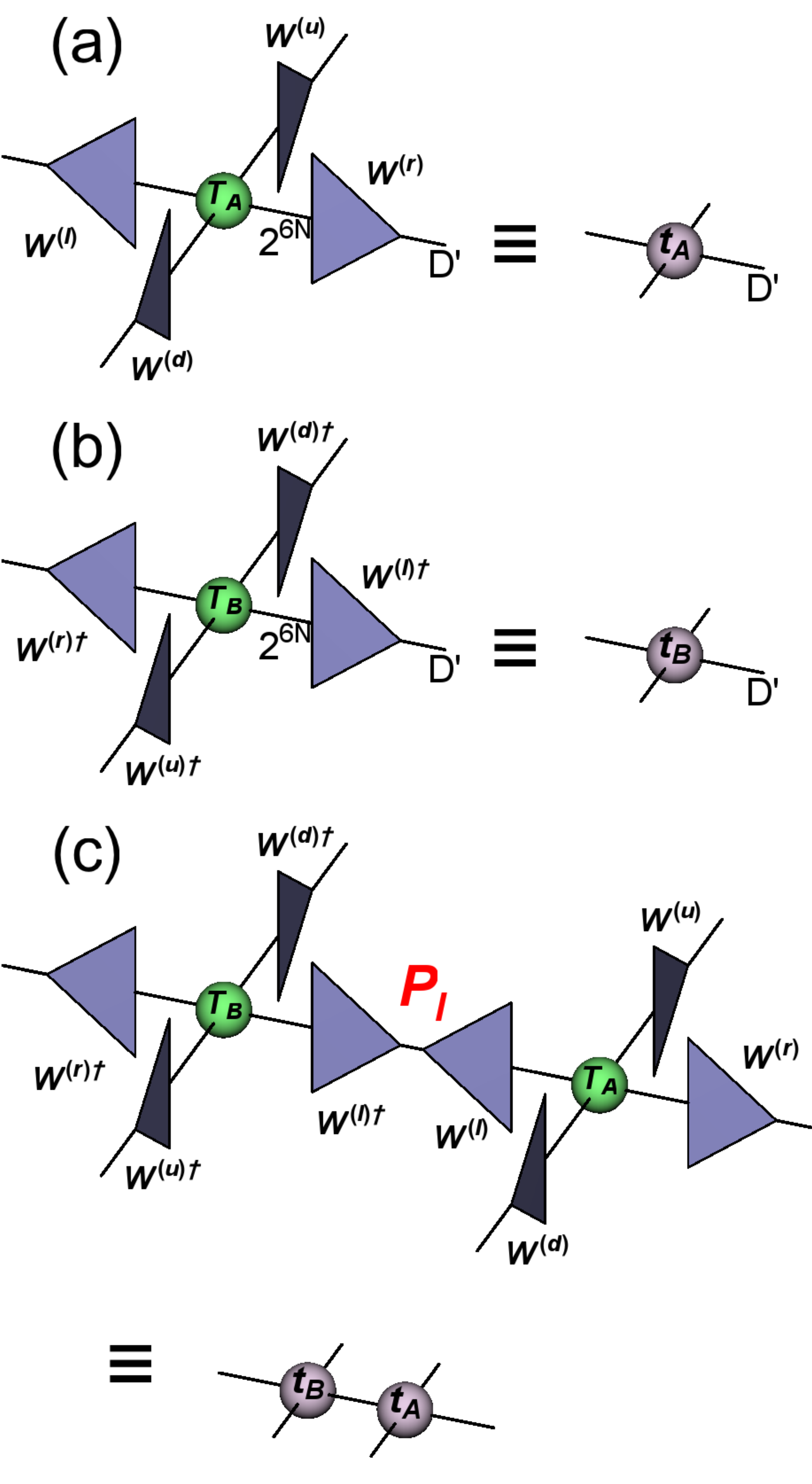}
\vspace{-0cm}
\caption{
In (a),
the transfer tensor $t_A$ is obtained from a tensor $T^A$ made of $N$ elementary tensors $T^A_0$, 
see Figs. \ref{fig:TTN} and \ref{fig:Tn}a. 
A bond dimension of $T_A$ is renormalized from huge $2^{6N}$ down to $D'$.
The renormalization along lattice bond $x$ is done by an isometry $W^{(x)}$, where $x=l,r,u,d$.
The huge isometry is a tree tensor network of small isometries $W_m^{(x)}$, 
compare Fig.~\ref{fig:TTN}.
In (b),
the transfer tensor $t_B$ has a similar structure as $t_A$. 
In (c), 
any contraction along a bond between   $t_A$ and  $t_B$, 
like in the partition function in Fig.~\ref{fig:Tn}b,
involves an orthogonal projection inserted between $T^A$ and $T^B$. 
Here the tensors  are contracted along their  common  $l$-bond   and 
the projection is $P_l=W^{(l)\dag}W^{(l)}$.
}
\label{fig:Iso}
\end{figure}
%%%%%%%%%%%%%%%%%%%%%%%%%%%%%%%%%%%%%%%%%%%%%%%%%%%%%%%%%%%%%%%%%%%%%%%%%%

%%%%%%%%%%%%%%%%%%%%%%%%%%%%%%%%%%%%%%%%%%%%%%%%%%%%%%%%%%%%%%%%%%%%%%%%%%%%
\begin{figure}[t!]
\vspace{-0cm}
\includegraphics[width=0.7\columnwidth,clip=true]{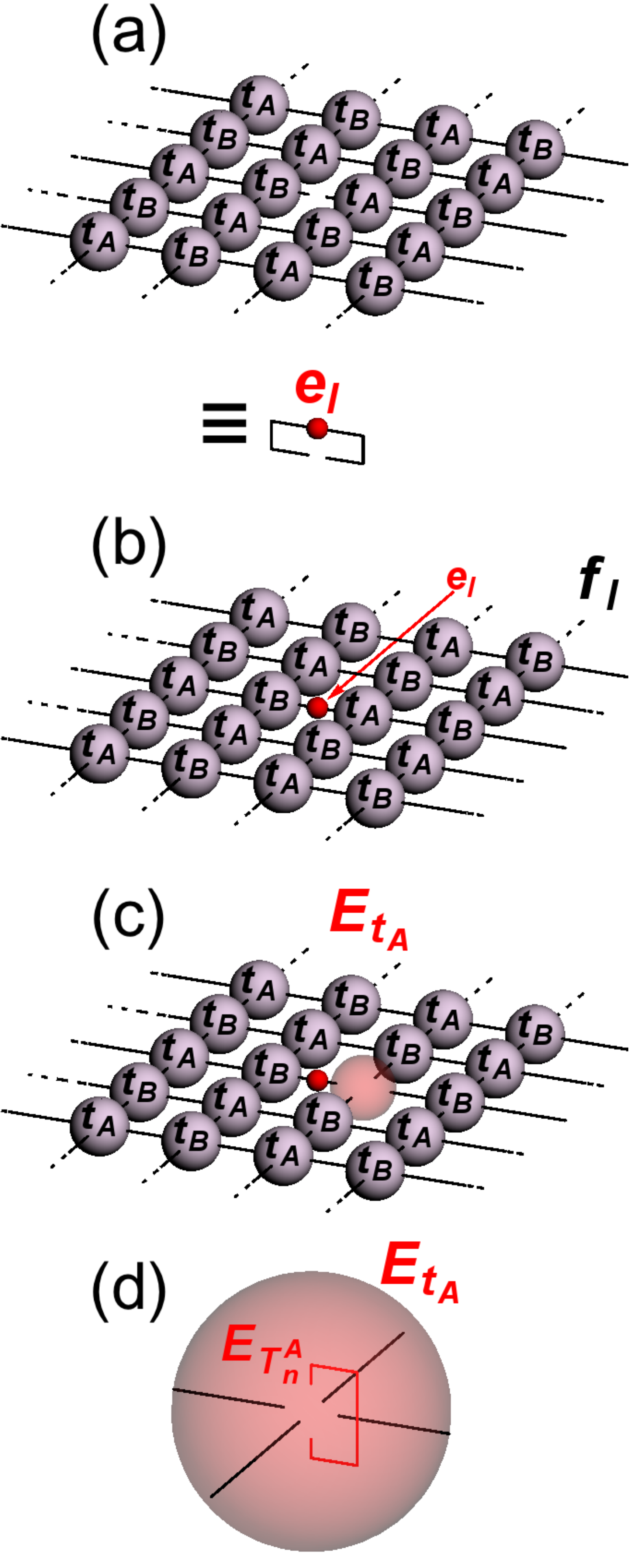}
\vspace{-0cm}
\caption{  
In (a),
a bond environment $e_l$ is obtained after cutting one of $l$-bonds in the partition 
function in Fig.~\ref{fig:Tn}c.
In (b),
the figure of merit $f_l$ in Eq.~\eqref{fl} is obtained by inserting a tensor representing
$e_l$ into the cut bond in panel (a).
In (c),
tensor environment $E_{t_A}$ for the tensor $t_A$ adjacent to the optimized $l$-bond is obtained 
by removing the adjacent $t_A$ from the figure of merit in panel (b).
In (d),
tensor environment $E_{T^A_n}$ for PEPO tensor $T^A_n$ is obtained from $E_{t_A}$.
}
\label{fig:Envt}
\end{figure}
%%%%%%%%%%%%%%%%%%%%%%%%%%%%%%%%%%%%%%%%%%%%%%%%%%%%%%%%%%%%%%%%%%%%%%%%%%%%

%%%%%%%%%%%%%%%%%%%%%%%%%%%%%%%%%%%%%%%%%%%%%%%%%%%%%%%%%%%%%%%%%%%%%%%%%%%%
\begin{figure}[t!]
\vspace{-0cm}
\includegraphics[width=0.99\columnwidth,clip=true]{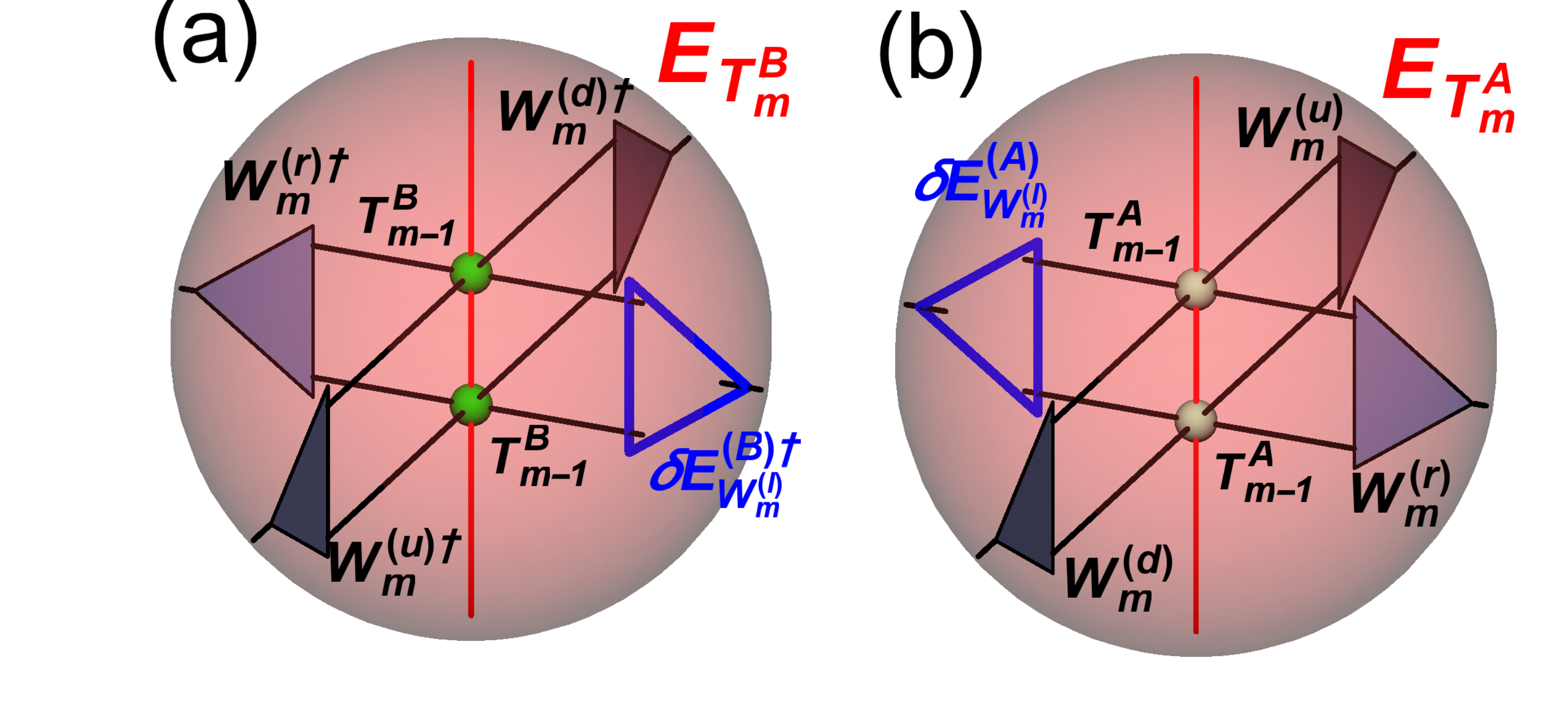}
\vspace{-0cm}
\caption{  
There are two inequivalent contributions to the tensor environment of isometry $W^{(l)}_m$:
$\delta E^{(B)}_{W^{(l)}_m}$ from sublattice $B$ in panel (a) and $\delta E^{(A)}_{W^{(l)}_m}$ from sublattice $A$ in panel (b).
They add up to form a full environment $E_{W^{(l)}_m}=\delta E^{(A)}_{W^{(l)}_m}+\delta E^{(B)}_{W^{(l)}_m}$.
}
\label{fig:EnvW}
\end{figure}
%%%%%%%%%%%%%%%%%%%%%%%%%%%%%%%%%%%%%%%%%%%%%%%%%%%%%%%%%%%%%%%%%%%%%%%%%%%% 

%%%%%%%%%%%%%%%%%%%%%%%%%%%%%%%%%%%%%%%%%%%%%%%%%%%%%%%%%%%%%%%%%%%%%%%%%%%% 
\subsection{Figure of merit}
\label{sec:FOM} 
%%%%%%%%%%%%%%%%%%%%%%%%%%%%%%%%%%%%%%%%%%%%%%%%%%%%%%%%%%%%%%%%%%%%%%%%%%%%

Figure \ref{fig:TTN} shows the net outcome after $3$ renormalization steps. 
There is a central column of $2^3$ elementary Trotter tensors $T^A_0$.
Their indices in the direction of bond $x$ are renormalized down to  the  bond 
dimension $D$ by $3$ layers of isometries $W^{(x)}_m$. The isometries 
form a tree tensor network that is also an isometry. A similar picture
could be drawn for the total of $n$ coarse-graining steps with a column
of $N=2^n$ elementary $T^A_0$ in the center and $n$ layers of isometries 
$W^{(x)}_m$ in each tree tensor network. Therefore, as shown in Fig.~\ref{fig:Iso}a, 
the transfer tensor $t_A$ is effectively obtained from a huge tensor $T^A$, where
$T^A$ is the central column of $N$ tensors $T^A_0$ traced over its (red) fermionic 
indices. The same is true for $t_B$, see Fig.~\ref{fig:Iso}b. Consequently, any 
contraction  of $t_A$  with $t_B$  corresponds to insertion of  an orthogonal projection  at every 
contracted bond index of  $T^A$  and $T^B$, see Fig.~\ref{fig:Iso}c. Such 
a projection is inserted, e.g., at every bond index of the partition function in Fig.~\ref{fig:Tn}b.
Were this projection an identity, the network in Fig.~\ref{fig:Tn}b would be the
exact partition function. Since it projects onto a small $D'$-dimensional subspace,
the isometries have to be optimized to minimize distortion of the partition function by projections. 
We expect that for large enough $D'$ each optimized projection becomes effectively an identity from the point of view of its tensor environment and the distortion becomes negligible.
We also expect that a more general non-orthogonal ansatz for the projection could accelerate convergence with $D'$ in some cases. This is not the case at half-filling (particle-hole symmetry) when the projections tensor environment is real-symmetric rendering the orthogonal ansatz general enough.
The more general projections are postponed to future work and here we continue with a construction of a figure of merit that has to be maximized in order to minimize the distortion. 

Figure \ref{fig:Envt}a defines $l$-bond environment $e_l$ which  is the partition function in Fig.~\ref{fig:Tn}c but with one of its $l$-bonds left uncontracted.  The uncontracted bond corresponds to ``cutting''  in half the projection $P_l=W^{(l)\dag}W^{(l)}$ introduced in Fig.~\ref{fig:Iso}c.
Therefore, the $D'\times D'$-dimensional bond environment $e_l$ can be rewritten as
\be 
e_l=W^{(l)}E_lW^{(l)\dag},
\label{el}
\ee
where we formally introduce a $2^{6N}\times2^{6N}$ unrenormalized bond environment $E_l$. 
 We want to minimize the distortion of $E_l$ by the isometries in Eq. \eqref{el} or, equivalently,
we want a projected environment $W^{(l)\dag}W^{(l)}E_lW^{(l)\dag}W^{(l)}=P_lE_lP_l$ to be as close
to the original $E_l$ as possible. The quality of this approximation is discussed in Appendix \ref{sec:spec}. To this end, we minimize the difference
\bea 
&& 
{\rm Tr}\,(P_lE_lP_l-E_l)^{\dag}(P_lE_lP_l-E_l)= \nonumber\\
&&
{\rm Tr}\,E_l^{\dag}E_l-{\rm Tr}\,E_l^{\dag}P_lE_lP_l= \nonumber\\
&&
{\rm Tr}\,E_l^{\dag}E_l-{\rm Tr}\,e_l^{\dag}e_l,
\label{EE}
\eea
where we used $P_lP_l=P_l=P_l^\dag$, $P_l=W^{(l)\dag}W^{(l)}$, and Eq. (\ref{el}). Since $E_l$
does not depend on the isometries $W^{(l)}$ at the considered $l$-bond, the first term
in (\ref{EE}) is constant with respect to these isometries and the actual figure of merit is
\be 
f_l~=~{\rm Tr}\,~e_l^\dag e_l.
\label{fl}
\ee 
It has to be maximized with respect to isometries $W^{(l)}_m$  at the considered $l$-bond.
Diagrammatically, $f_l$ can be obtained after inserting a tensor representing $e_l^{\dag}$ into 
the uncontracted bond of the network representing $e_l$, as shown in Fig.~\ref{fig:Envt}b. 

Due to translational invariance, the same isometries $W^{(l)}_m$ appear at every
$l$-bond in the infinite tensor network representing the partition function. They are optimized iteratively: once the $W^{(l)}_m$
 at the considered $l$-bond are optimized, the optimized $W^{(l)}_m$ are applied at all
equivalent $l$-bonds of the  partition function. The iteration is repeated until self-consistency
between different $l$-bonds is achieved. Similar figure of merits are simultaneously maximized for the other
three bonds: $f_r$, $f_u$, and $f_d$. The optimization is iterated until self-consistency between
different types of bonds is also achieved. The self-consistency is controlled by monitoring two-site reduced density matrices. In this work we demand their convergence up to $6$~decimal points.
 
%%%%%%%%%%%%%%%%%%%%%%%%%%%%%%%%%%%%%%%%%%%%%%%%%%%%%%%%%%%%%%%%%%%%%%%%%%%% 
\subsection{Variational optimization}
\label{sec:Var} 
%%%%%%%%%%%%%%%%%%%%%%%%%%%%%%%%%%%%%%%%%%%%%%%%%%%%%%%%%%%%%%%%%%%%%%%%%%%%

A preparatory step for variational optimization of isometries is calculation of the bond environment $e_l$ in Fig.~\ref{fig:Envt}a. This infinite network cannot be contracted exactly, but its accurate approximation, that can be improved in a systematic way, is obtained with the corner transfer matrix method (CTM) \cite{CTM}. We described the modification of the CTM algorithm of Ref.~\onlinecite{CorbozQR}, which we employ here, in Appendix \ref{sec:CTM}. Similar calculations are done in parallel to obtain $e_r$, $e_u$ and $e_d$. 

In order to optimize $f_l$ with respect to an isometry $W^{(l)}_m$ at the considered $l$-bond, 
we need a derivative of $f_l$ with respect to this isometry.
Since $f_l$ is a composed function, 
we calculate first derivatives of $f_l$ with respect to the two transfer tensors, $t_A$ and $t_B$, adjacent to the $l$-bond. 
The derivatives are proportional to the tensor environments $E_{t_A}$ and $E_{t_B}$ 
that are obtained by removing from $f_l$ the adjacent $t_A$ and $t_B$, respectively.
The construction of $E_{t_A}$ is shown in Fig.~\ref{fig:Envt}c. 
$E_{t_B}$ is constructed in an analogous way.

With $E_{t_A}$ at hand, 
it is easy to construct an environment $E_{T^A_n}$ of the PEPO tensor $T^A_n$, 
see Fig.~\ref{fig:Envt}d.
$E_{T^A_n}$ is proportional to a derivative of $f_l$ with respect to the $T^A_n$ adjacent to the
$l$-bond that we want to optimize.
A similar step from $E_{t_B}$ to $E_{T^B_n}$ is done in parallel. 
With both $E_{T^A_n}$ and $E_{T^B_n}$ at hand
a down optimization sweep for isometries $W^{(l)}_m$ can be initialized.

%%%%%%%%%%%%%%%%%%%%%%%%%%%%%%%%%%%%%%%%%%%%%%%%%%%%%%%%%%%%%%%%%%%%%%%%%%%%
\begin{figure}[t!]
\vspace{-0cm}
\includegraphics[width=0.65\columnwidth,clip=true]{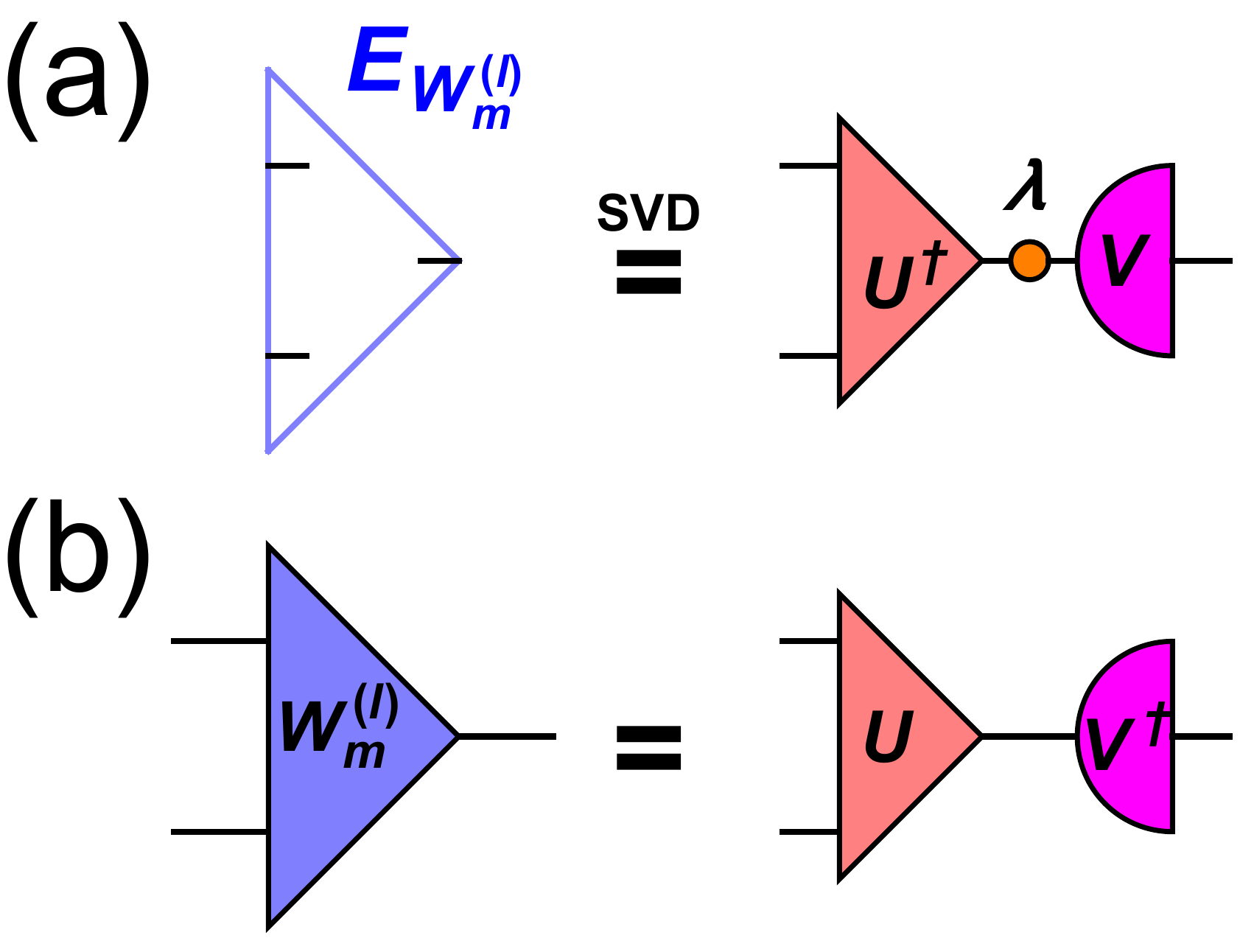}
\vspace{-0cm}
\caption{  
An update of isometry $W_m^{(l)}$, where the figure of merit 
$f_l \propto \textrm{Tr}\,  W_m^{(l)} E_{W^{(l)}_m}$.
In (a),
the environment is subject to singular value decomposition: 
$E_{W^{(l)}_m}=V\lambda U^\dag$.
In (b), 
the isometry is updated as $W_m^{(l)}=UV^\dag$.
}
\label{fig:svdEnvW}
\end{figure}
%%%%%%%%%%%%%%%%%%%%%%%%%%%%%%%%%%%%%%%%%%%%%%%%%%%%%%%%%%%%%%%%%%%%%%%%%%%% 

From, respectively, $E_{T^A_n}$ and $E_{T^B_n}$ two contributions, 
$\delta E^{(A)}_{W^{(l)}_n}$ and $\delta E^{(B)}_{W^{(l)}_n}$, to the environment $E_{W^{(l)}_n}$ are obtained
as presented in Fig.~\ref{fig:EnvW}.
They add up to a full environment, $E_{W^{(l)}_n}=\delta E^{(A)}_{W^{(l)}_n}+\delta E^{(B)}_{W^{(l)}_n}$. 
The environment $E_{W^{(l)}_n}$ is used immediately to update $W^{(l)}_n$ by singular value decomposition,
see Fig.~\ref{fig:svdEnvW}.

\begin{figure}[t!]
\vspace{-0cm}
\includegraphics[width=0.99\columnwidth,clip=true]{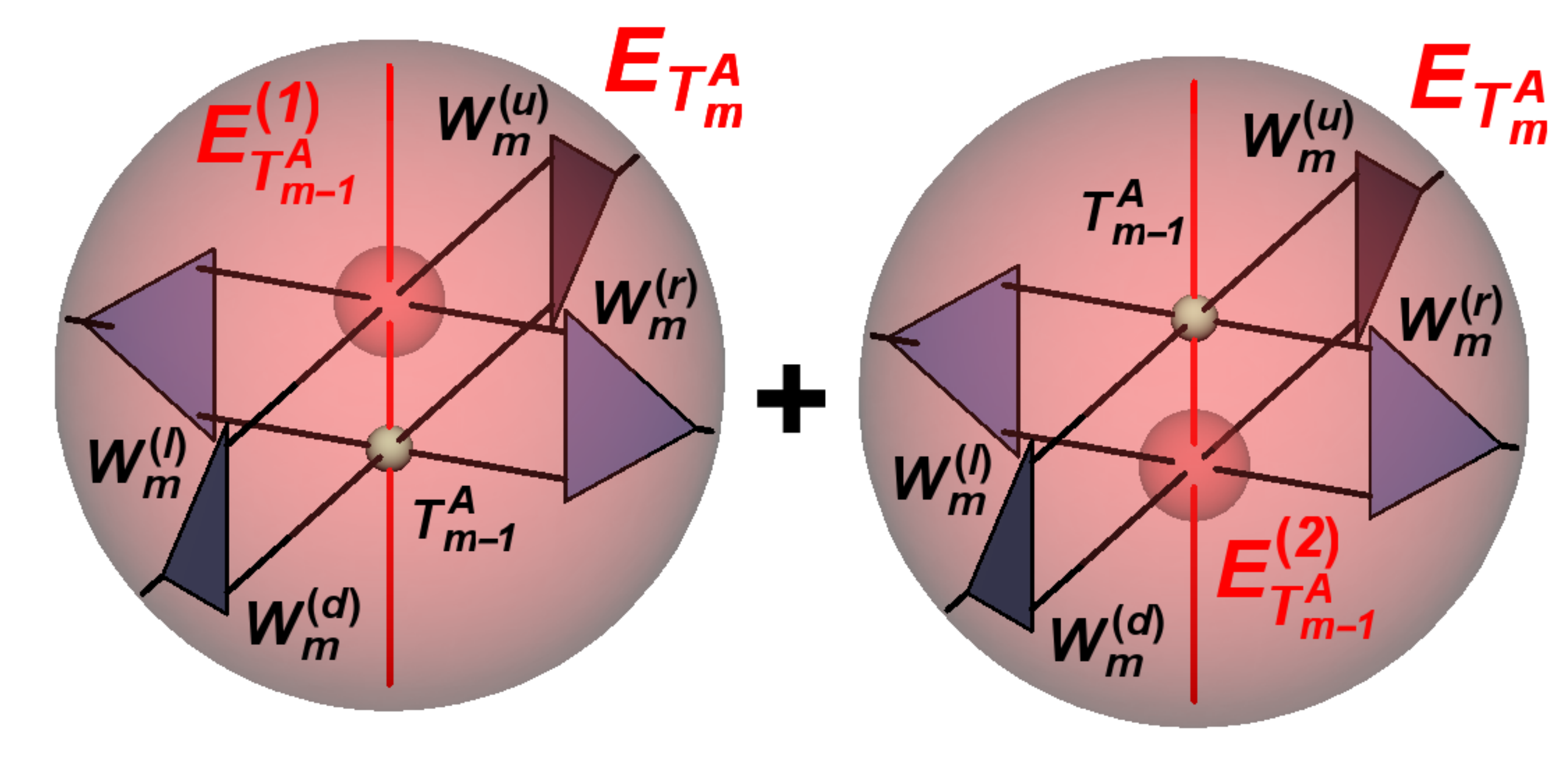}
\vspace{-0cm}
\caption{  
There are two inequivalent contributions to the environment of $T^A_{m-1}$:
$E^{(1)}_{T^A_{m-1}}$ and $E^{(2)}_{T^A_{m-1}}$, which add up to form the full environment
$E_{T^A_{m-1}}=E^{(1)}_{T^A_{m-1}}+E^{(2)}_{T^A_{m-1}}$.
}
\label{fig:ETm}
\end{figure}

With the updated $W^{(l)}_n$ the first step down in the TTN hierarchy can be done from $E_{T^{A,B}_n}$ to $E_{T^{A,B}_{n-1}}$, 
see Fig.~\ref{fig:ETm}.
Then, new environments $E_{W^{(l)}_{n-1}}$ are obtained from $E_{T^{A,B}_{n-1}}$ and used immediately to update the isometries $W^{(l)}_{n-1}$. 
The same procedure is repeated all the way down to $W^{(l)}_1$ whose update completes the down-sweep.

Once $W^{(l)}_1$ and $T_1^{A,B}$ are updated, an up optimization sweep begins. In has $n$ steps.
In the $m$-th step the tensors $T^{A,B}_{m-1}$ and the environments $E_{T_m^{A,B}}$ --- calculated before during the down-sweep --- 
are contracted to obtain the environments $E_{W^{(l)}_m}$ and update the isometries $W^{(l)}_m$, see Fig.~\ref{fig:EnvW}. 
The updated $W^{(l)}_m$ are used to coarse-grain $T^{A,B}_{m-1}\times T^{A,B}_{m-1}\to T^{A,B}_m$, see Figs. \ref{fig:Tm}b,c. 
This basic step is repeated all the way up to $T_n$.  

%%%%%%%%%%%%%%%%%%%%%%%%%%%%%%%%%%%%%%%%%%%%%%%%%%%%%%%%%%%%%%%%%%%%%%%%%%%%
%%%%%%%%%%%%%%%%%%%%%%%%%%%%%%%%%%%%%%%%%%%%%%%%%%%%%%%%%%%%%%%%%%%%%%%%%%%%  

The up-sweep completes one optimization loop consisting of three stages:
\bi
\item the CTM procedure:
$$ 
T^{A,B}_n\,\stackrel{\rm CTM}{\longrightarrow}\, E_{t_{A,B}}\to\, E_{T^{A,B}_n};
$$
\item the down-sweep:
$$
E_{T^{A,B}_n}\to\, E_{W^{(l)}_n}\to\, E_{T^{A,B}_{n-1}}\to\dots\to\, E_{T^{A,B}_1}\to\, E_{W^{(l)}_1};
$$
\item the up-sweep:
$$
T^{A,B}_0\to\, T^{A,B}_1\to E_{W^{(l)}_2}\to T^{A,B}_{2} \to\dots\to\,  E_{W^{(l)}_n}\to \, T^{A,B}_n.
$$
\ei
Here each $E_{W^{(l)}_m}$ is used immediately to update $W^{(l)}_m$. 
Similar optimization sweeps are done in parallel for $W^{(r)}_m$, $W^{(u)}_m$ and $W^{(d)}_m$.
The loop is repeated until convergence. 
The dominant numerical cost of the procedures in this section scales like $D^8$. 
Typically, it is subleading as compared to the cost of CTM. 

Having outlined the algorithm we can now proceed with the benchmark results for the Hubbard model.

%%%%%%%%%%%%%%%%%%%%%%%%%%%%%%%%%%%%%%%%%%%%%%%%%%%%%%%%%%%%%%%%%%%%%%%%%%%%%%%
\section{Benchmark results}
\label{sec:results}
%%%%%%%%%%%%%%%%%%%%%%%%%%%%%%%%%%%%%%%%%%%%%%%%%%%%%%%%%%%%%%%%%%%%%%%%%%%%%%%%

%%%%%%%%%%%%%%%%%%%%%%%%%%%%%%%%%%%%%%%%%%%%%%%%%%%%%%%%%%%%%%%%%%%%%%%%%%%%%%%
\subsection{Non-interacting case}
\label{sec:FermiSea}
%%%%%%%%%%%%%%%%%%%%%%%%%%%%%%%%%%%%%%%%%%%%%%%%%%%%%%%%%%%%%%%%%%%%%%%%%%%%%%%%

%%%%%%%%%%%%%%%%%%%%%%%%%%%%%%%%%%%%%%%%%%%%%%%%%%%%%%%%%%%%%%%%%%%%%%%%%%%%
\begin{figure}[t!]
\vspace{-0cm}
\includegraphics[width=0.99\columnwidth,clip=true]{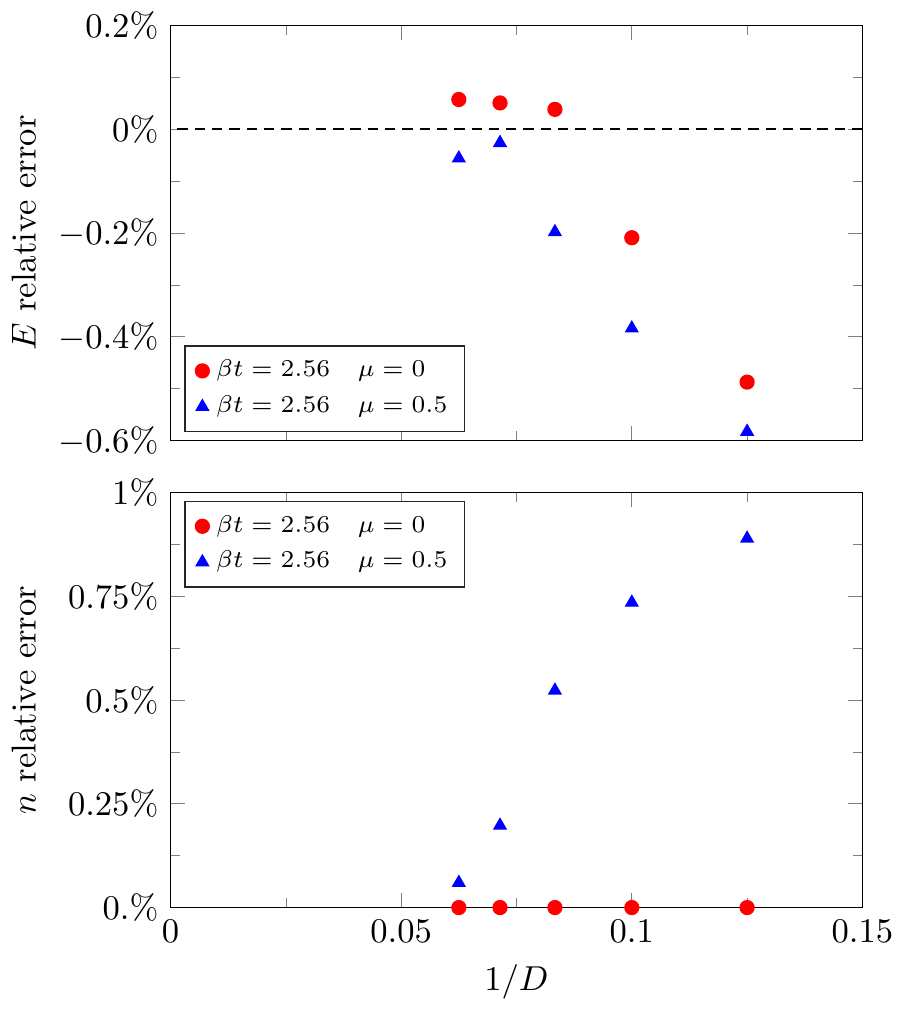}
\vspace{-0cm}
\caption{  
Relative errors of energy $E$ and average fermion density $n$ for non-interacting spinless fermions at temperature $\beta t = 2.56$ and two values of chemical potential as a function of the bond dimension $8\leq D \leq 16$.
}
\label{fig:resFS}
\end{figure}
%%%%%%%%%%%%%%%%%%%%%%%%%%%%%%%%%%%%%%%%%%%%%%%%%%%%%%%%%%%%%%%%%%%%%%%%%%%%  

We begin testing our algorithm in an exactly solvable version of the Hamiltonian \eqref{H}
without the interaction term, $U=0$, and with only one specie of spin-less fermions. 
The thermal state is the finite-temperature Fermi sea which, while being a trivial product state in 
the quasi-momentum basis,  for large $\beta$  is strongly entangled in the position representation and therefore pose a challenge for tensor networks.  

Numerical results for the inverse temperature $\beta t=2.56$ are collected in Fig.~\ref{fig:resFS}.
We show results for the chemical potential $\mu=0$ (the half-filled Fermi sea) and $\mu=0.5$. 
We present relative errors of energy $E$ and fermion density $n$. $E$ is defined as energy per site where we neglect the contribution related to the chemical potential -- this is done in order to match  convention of  Ref.~\onlinecite{Hubbardreview}. We see systematic improvement with growing bond dimension $D$, with the relative error below $0.1\%$ for the largest $D=16$ used here. 

%%%%%%%%%%%%%%%%%%%%%%%%%%%%%%%%%%%%%%%%%%%%%%%%%%%%%%%%%%%%%%%%%%%%%%%%%%%%%%%
\subsection{Strongly correlated case}
\label{sec:U}
%%%%%%%%%%%%%%%%%%%%%%%%%%%%%%%%%%%%%%%%%%%%%%%%%%%%%%%%%%%%%%%%%%%%%%%%%%%%%%%%

%%%%%%%%%%%%%%%%%%%%%%%%%%%%%%%%%%%%%%%%%%%%%%%%%%%%%%%%%%%%%%%%%%%%%%%%%%%%
\begin{figure}[t!]
\vspace{-0cm}
\includegraphics[width=0.99\columnwidth,clip=true]{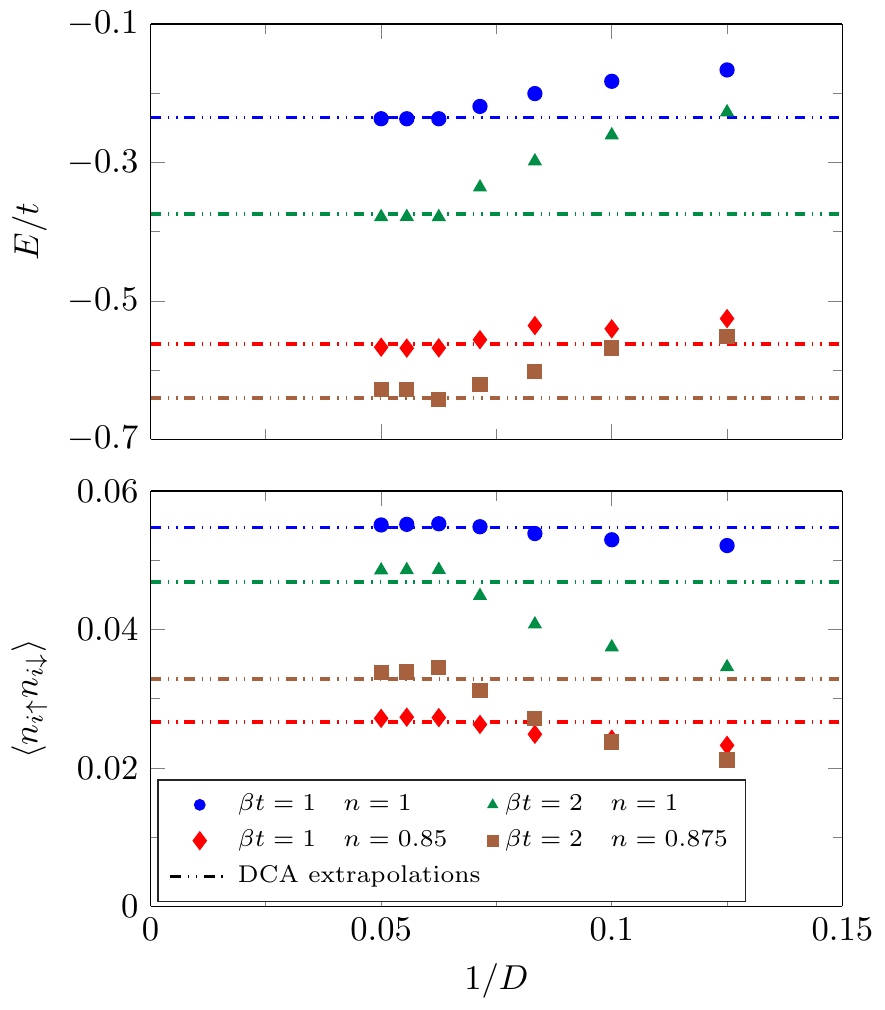}
\vspace{-0cm}
\caption{  
Points show energy $E$ and double occupancy $\langle n_{i\uparrow} n_{i\downarrow} \rangle$  as a function of the bond dimension $8 \leq D \le 20$ obtained for the Hubbard model in Eq.~\eqref{H} for $U/t=8$. Circles, triangles, squares and diamonds mark different temperatures or electron densities. Errors  of  electron density $n$ for the given bond dimension $D$, related to scanning for the correct value of the chemical potential $\mu$, are smaller than 0.001.
For comparison, lines show corresponding results from dynamical cluster approximation calculations extrapolated with the cluster size sent to infinity \cite{Hubbardreview,LeBla13}. Values of $\mu$ for the points away from $n=1$ are shown in Tab.~\ref{tab:mu}. 
}
\label{fig:resHubb}
\end{figure}
%%%%%%%%%%%%%%%%%%%%%%%%%%%%%%%%%%%%%%%%%%%%%%%%%%%%%%%%%%%%%%%%%%%%%%%%%%%%  

Finally, to benchmark our algorithm we simulate strongly correlated case of  $U/t=8$. We collect  the results at inverse temperatures $\beta t= 1$ and $\beta t = 2$,
and for electron densities $n=1$, $0.875$, and $0.85$ in Fig.~\ref{fig:resHubb}. We compare them with  DCA results which are obtained by extrapolating the cluster size to infinity \cite{Hubbardreview,LeBla13}. The DCA results at $n=1$, $\beta t = 1,2$ and  $n=0.85$, $\beta t = 1$  were shown to be consistent with NLCE and DQMC results\cite{Hubbardreview,LeBla13}, so they can be treated as a benchmark. 
With the increasing bond dimension $D$ our results approach  those obtained by DCA.  For the largest simulated $D=20$ the energies $E$ (the double occupancies $\langle n_{i,\uparrow} n_{i,\downarrow} \rangle$) agree well with DCA as relative discrepancies are smaller than $1.9\%$ ($3.6\%$). The values of chemical potential used in the simulations are listed below in Table.~\ref{tab:mu}.

\begin{table}[hbt!]
\begin{tabular}{|c|c|c|}
\hline
$D$ & $\beta=1,\, n = 0.85$ & $\beta=2,\, n=0.875$ \\
\hline \hline
8 & 2.254 & 2.064 \\\hline
10 & 2.302 & 2.085 \\\hline
12 & 2.274 & 2.099 \\\hline
14 & 2.334 & 2.148 \\\hline
16 & 2.351 & 2.259 \\\hline
18 & 2.356 & 2.232 \\\hline
20 & 2.349 & 2.212 \\
\hline
\end{tabular}
\caption{
Values of chemical potential $\mu' = (\mu + U/2) /t$  for points from Fig.~\ref{fig:resHubb}. We shift the values of the chemical potential above by $U/2t = 4$ to match the common convention.} 
\label{tab:mu}
\end{table}

%%%%%%%%%%%%%%%%%%%%%%%%%%%%%%%%%%%%%%%%%%%%%%%%%%%%%%%%%%%%%%%%%%%%%%%%%%%%%%%
\subsection{Numerical details}
\label{sec:details}
%%%%%%%%%%%%%%%%%%%%%%%%%%%%%%%%%%%%%%%%%%%%%%%%%%%%%%%%%%%%%%%%%%%%%%%%%%%%%%%%

All simulations were made with MATLAB using  {\it ncon} tensor network contractor \cite{encon} and representing tensors with real numbers. The results for $U/t=8$ were obtained using 3 desktops during 3 weeks.  

Isometries $W_{(x)}^m$ for $U/t=8$ were initialized by  local tensor update. It was checked for the non-interacting case that such initialization leads to the same results as initialization by random isometries.

  The results were already converged  in the environmental bond dimension $M$ (see Appendix \ref{sec:CTM}) for  $M=30$  in the case of $U/t=8$ and for $M=50$ in the non-interacting case. They are also converged in the small Trotter step $d\beta$ for $d\beta t \le 0.001$.
  
%%%%%%%%%%%%%%%%%%%%%%%%%%%%%%%%%%%%%%%%%%%%%%%%%%%%%%%%%%%%%%%%%%%%%%%%%%%%%%
\section{Conclusion}
\label{sec:conclusion}
%%%%%%%%%%%%%%%%%%%%%%%%%%%%%%%%%%%%%%%%%%%%%%%%%%%%%%%%%%%%%%%%%%%%%%%%%%%%%%

In this paper we presented a fermionic version of the algorithm that was introduced and tested in the 2D quantum Ising model in Ref.~\onlinecite{var} and applied to the quantum compass model in Ref.~\onlinecite{ourcompass}. The Gibbs operator $e^{-\beta H}$ for a 2D lattice system can be represented by a 3D tensor network, where the third dimension is the imaginary time (inverse temperature) $\beta$. Thanks to the area law for thermal states, coarse-graining the 3D network along $\beta$ results in an accurate 2D projected entangled-pair operator (PEPO) with a finite bond dimension $D'$. The coarse-graining is performed by a tree tensor network (TTN) of  isometries that are optimized variationally to maximize the accuracy of the final PEPO. 

Here the algorithm was applied to the fermionic Hubbard model on an infinite square lattice. Results were obtained at $\beta t=1$ and $\beta t=2$ for electronic densities $n=1$, $0.85,$ and $0.875$. In this regime different numerical  methods (DCA, NLCE, DQMC) yield mutually consistent results that are considered to be a benchmark test for any new algorithm \cite{Hubbardreview}. The results in this paper demonstrate that our method meets these standards.

The fermionic algorithm presented here is largely a straightforward generalization of the algorithm developed for spin models \cite{var,ourcompass}. The necessary modification are the -- by now standard -- swap gates that account for the anti-commuting character of the degrees of freedom. A novel algorithmic feature of this paper is a numerically efficient modification of corner matrix renormalization algorithm that, however, is a general development not addressing any specific needs of the strongly correlated fermions. 
While the current approach is yielding  promising results there is still room for further algorithmic improvements,  for instance in the design of dimensional reduction scheme which is limited to orthogonal projections in the current approach.

%%%%%%%%%%%%%%%%%%%%%%%%%%%%%%%%%%%%%%%%%%%%%%%%%%%%%%%%%%%%%%%%%%%%%%%%%%%%%%
\acknowledgments
%%%%%%%%%%%%%%%%%%%%%%%%%%%%%%%%%%%%%%%%%%%%%%%%%%%%%%%%%%%%%%%%%%%%%%%%%%%%%%

We kindly acknowledge support by Narodowe Centrum Nauki (National Science Center) under Project No. 2013/09/B/ST3/01603.
The work of P.C. on his Ph.D. thesis was supported by Narodowe Centrum Nauki under Project No. 2015/16/T/ST3/00502.

%%%%%%%%%%%%%%%%%%%%%%%%%%%%%%%%%%%%%%%%%%%%%%%%%%%%%%%%%%%%%%%%%%%%%%%%%%%%%%
\appendix
%%%%%%%%%%%%%%%%%%%%%%%%%%%%%%%%%%%%%%%%%%%%%%%%%%%%%%%%%%%%%%%%%%%%%%%%%%%%%%

%%%%%%%%%%%%%%%%%%%%%%%%%%%%%%%%%%%%%%%%%%%%%%%%%%%%%%%%%%%%%%%%%%%%%%%%%%%%%%%
\section{Corner transfer matrix renormalization}
\label{sec:CTM}
%%%%%%%%%%%%%%%%%%%%%%%%%%%%%%%%%%%%%%%%%%%%%%%%%%%%%%%%%%%%%%%%%%%%%%%%%%%%%%%%

\begin{figure}[t!]
%\vspace{-0cm}
\includegraphics[width=0.90\columnwidth,clip=true]{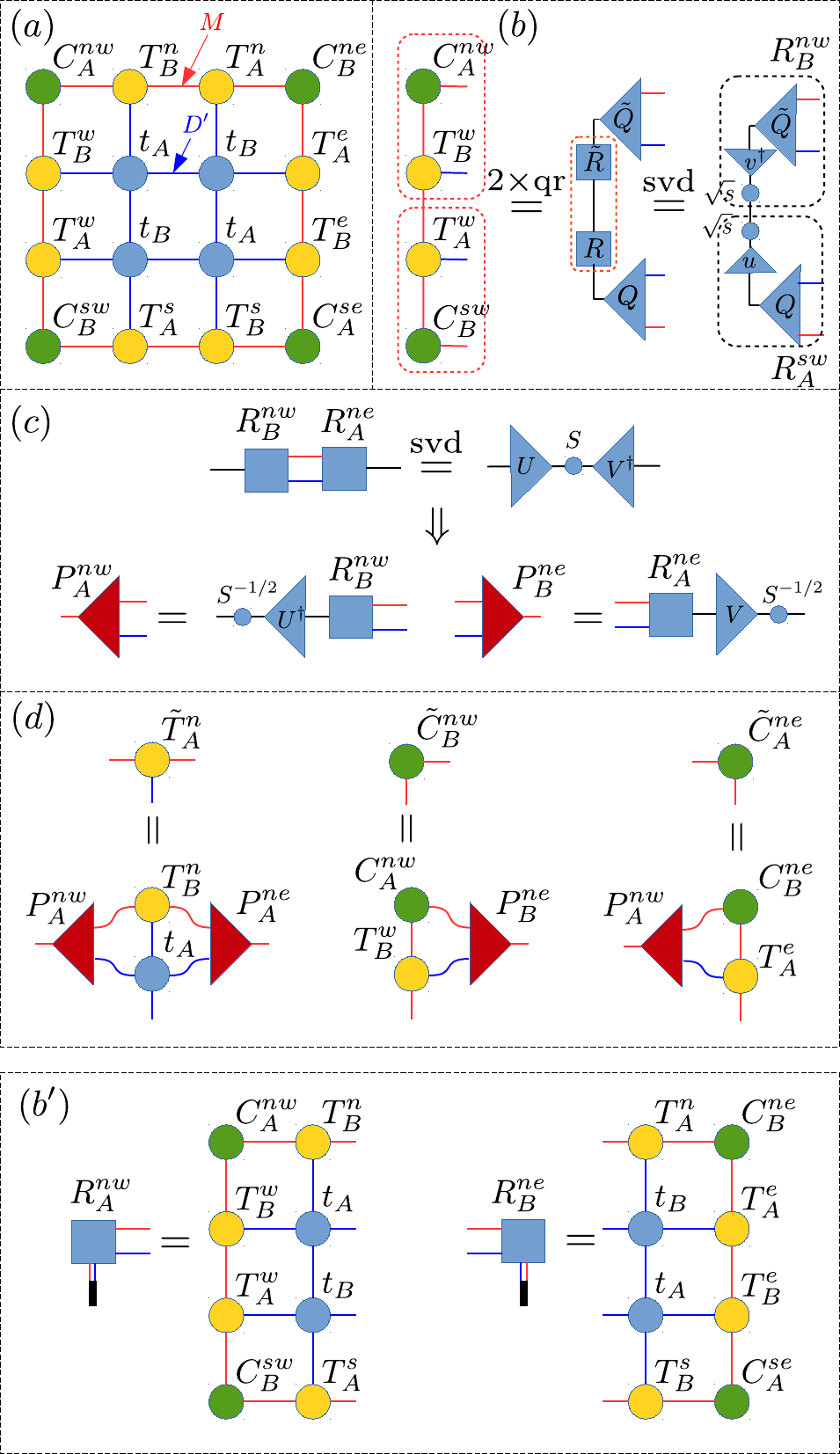}
\caption{
Details of a single up move in the corner transfer matrix algorithm. (a)  A block of $2 \times 2$ $t_{A(B)}$ tensors on a chessboard, where its environment is represented by environment tensors $C$ and $T$. (b) West half-plane is divided using SVD with singular values split symmetrically between the two parts. Performing QR decomposition in the middle step increases efficiency. The same is done for east half-plane (not shown). (c) SVD of $R^{nw}_{A}R^{ne}_B$ (and similarly $R^{nw}_{B}R^{ne}_A$) is performed to find the projections $P$, which, in (d), are used to truncate the enlarged edge and corner tensors. Step (b) replaces (b$'$) from Ref.~\onlinecite{CorbozQR}. If (b$'$) is used instead, additionally, only $M$ largest singular values are kept to calculate projections in (c). }
\label{fig:CTMfigure}
\end{figure}

The numerically most expensive part of the algorithm is computation of the environment of a single bond in Fig.~\ref{fig:Envt}a.
We employ the modified version of the corner-transfer-matrix renormalization (CTM) algorithm described in Ref.~\onlinecite{CorbozQR}. The alternation boils down to the strategy of finding the projections which renormalize  enlarged CTM corner and edge tensors, which, while being significantly less expensive, turns out to be numerically almost as stable as the original one. 
For completeness, in Fig.~\ref{fig:CTMfigure} we illustrate all the steps making up a single CTM update for a network with unit cell on a chessboard used in Fig.~\ref{fig:Envt}a.

The CTM algorithm computes corner $C^{nw},C^{ne},C^{se},C^{sw}$ and edge $T^n,T^e,T^s,T^w$ tensors which encode the infinite two-dimensional  network surrounding bulk tensors $t$, as shown in Fig.~\ref{fig:CTMfigure}(a), where the subscript $_{A(B)}$ labels the position in the unit cell. We refer to Ref.~\onlinecite{CorbozQR} for the details of the method and focus here on differences. 

Fig.~\ref{fig:CTMfigure}(b)-(d) shows how to perform the up (north) move.  In Fig.~\ref{fig:CTMfigure}b we divide the enlarged environment corresponding to west half-plane into the north-west $R^{nw}$ and the south-west $R^{sw}$ parts using SVD where singular values are symmetrically split between the two parts. For a choice of effective half-plane in Fig.~\ref{fig:CTMfigure}(b) we exploit that it is not full-rank: by computing the QR decomposition as a middle step, we are left with SVD of $M\times M$ matrix, instead of not-full-rank $D' M\times D' M$ matrix. Similarly, we divide east half-plane into $R^{ne}$ and $R^{se}$ (not shown). This substitutes the step (b') used in Ref.~\onlinecite{CorbozQR}.
Matrices $R^{nw}$ and $R^{ne}$ are then used in (c) to compute the projections $P^{nw}$ and $P^{ne}$. Finally, in (d) we update the north edge and corner tensors by using the projections $P$ to truncate the enlarged old ones. This completes single up move.  The down and up moves  on both sublattices $_{A}$ and $_B$ are  all performed simultaneously.  The procedure is then run iteratively performing up-down and left-right moves until convergence.

We note that in step (b) $R^{nw}$ etc. are $M \times D' M$ matrices, so one only needs to find SVD of $M \times M$ matrix to calculate projections in (c), comparing to SVD of $D' M \times D' M$ matrix if (b') is used instead. This way, for the range of $D' = D^2 \le 20^2$ used in this article the numerically most expensive part of the modified CTM update becomes the renormalization of the edge tensor in step (d). From our experience the above procedure is almost as stable as the original one and within the precision of the main algorithm converges to the same values of local observables. What is more, the singular values obtained in step (c) are typically by a factor of square root larger then when (b$'$) is used instead, which suggest better numerical stability of calculating the projections $P$ (where the singular values are inverted) -- we however don't see that this is of relevance within the required precision.

The biggest disadvantage of using step (b) is that it does not allow to systematically increase the bond-dimension $M$ of corner and edge tensors in a single CTM step. That is why we find it optimal to combine the two strategies and perform a single expensive move using (b$'$) after every $10-20$ moves using (b), where we observe that the two strategies are complementary in a sense that typically the procedure is monotonically converging. While greatly reducing the time needed for CTM to converge this provides systematic way of increasing the CTM bond dimension $M$ to converged in.

The procedure can be understood as analogical to the truncation in infinite Matrix Product State (iMPS) \cite{Orus08} -- in principle biorthogonal \cite{Huang12} as the Matrix Product Operator constructed from row or column of $t$ tensors in principle is not hermitian --  where the (enlarged) edge tensors play the role of the MPS tensors to be truncated, motivating the form of projections used in (c) and (d). The difference from the MPS procedure is that there the $R$ matrices would correspond to (square roots) of the left and right dominant eigenvectors of MPS transfer matrix which are calculated at every iteration. Here, however, they are found as a result of iterative procedure as CTM is converging. We indeed observe that while CTM procedure is converging, the combined pairs of neighboring corners $C$ are converging to the dominant eigenvectors of the corresponding transfer matrices constructed from opposing edge tensors $T$.

Finally, we note that in Ref.~\onlinecite{CorbozQR} QR decompositions of the $R$ matrices in (b$'$) are computed and only the upper-triangular parts of those decompositions are retained to calculate the projections in (c). On paper, this intermediate step does not change the projections, we see, however, that it indeed increases numerical stability of the CTM algorithm. We perform similar middle step on $R^{nw}$, $R^{ne}$ obtained in (b), which, for clarity of presentation, is not explicitly shown in Fig.~\ref{fig:CTMfigure}.

%%%%%%%%%%%%%%%%%%%%%%%%%%%%%%%%%%%%%%%%%%%%%%%%%%%%%%%%%%%%%%%%%%%%%%%%%%%%%%%%%%%%%%%
\section{Comparison with direct imaginary time evolution.}
\label{sec:imtev}
%%%%%%%%%%%%%%%%%%%%%%%%%%%%%%%%%%%%%%%%%%%%%%%%%%%%%%%%%%%%%%%%%%%%%%%%%%%%
\begin{figure}[h!]
\vspace{-0cm}
\includegraphics[width=0.99\columnwidth,clip=true]{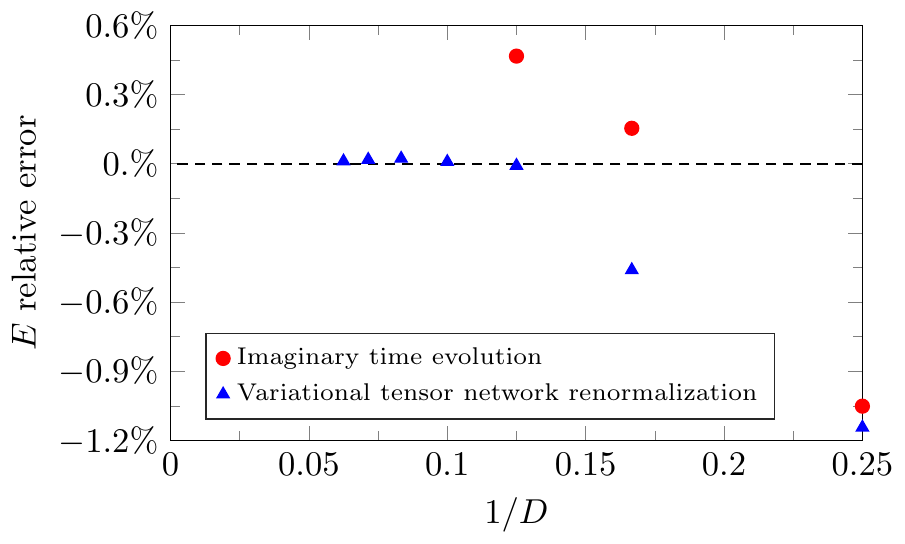}
\vspace{-0cm}
\caption{  
Comparison of  energy $E$ convergence with increasing $D$ for variational tensor network renormalization and imaginary  time evolution \cite{evolution}. Points represent $E$   as a function of the bond dimension ($4 \leq D \le 16$  for the variational approach and $4 \leq D \le 8$ for the imaginary time evolution)  for non-interacting spinless fermions at temperature $\beta t = 1.28$ and chemical potential $\mu = 0$. Data for the imaginary evolution are taken from Ref.~\onlinecite{evolution}.
}
\label{fig:imtev}
\end{figure}
Comprehensive comparison of the variational tensor network renormalization with the imaginary time evolution approach\cite{Czarniks,evolution,self} is presented in Ref.~\onlinecite{var}. There, two of us argued for superiority of the variational approach implying that in that case both larger values of the bond dimension $D$  should be accessible and faster convergence of observables in $D$ can be expected. 
Those arguments were corroborated by significantly better accuracy of the results obtained with the variational procedure in the ferromagnetic phase of 2D quantum Ising model close to the critical point. A fundamental problem faced while simulating imaginary time evolution in that case is related to crossing the critical point between the high temperature paramagnetic and low temperature ferromagnetic phases, which is however circumvented in the variational approach directly targeting the final state. Here, in Fig.~\ref{fig:imtev}, we compare the results obtained with both procedures for non-interacting spinless fermions for relatively high temperature far from the critical point at zero temperature showing that even in such a situation both larger possible values of $D$ and faster convergence of observables in $D$ are achieved by the variational tensor network renormalization.  
%%%%%%%%%%%%%%%%%%%%%%%%%%%%%%%%%%%%%%%%%%%%%%%%%%%%%%%%%%%%%%%%%%%%%%%%%%%%  

%%%%%%%%%%%%%%%%%%%%%%%%%%%%%%%%%%%%%%%%%%%%%%%%%%%%%%%%%%%%%%%%%%%%%%%%%%%%%%%%%%%%%%%
\section{Spectrum of the bond environment}
\label{sec:spec}
%%%%%%%%%%%%%%%%%%%%%%%%%%%%%%%%%%%%%%%%%%%%%%%%%%%%%%%%%%%%%%%%%%%%%%%%%%%%%%%%%%%%%%%

\begin{figure}[h!]
\vspace{-0cm}
\includegraphics[width=0.99\columnwidth,clip=true]{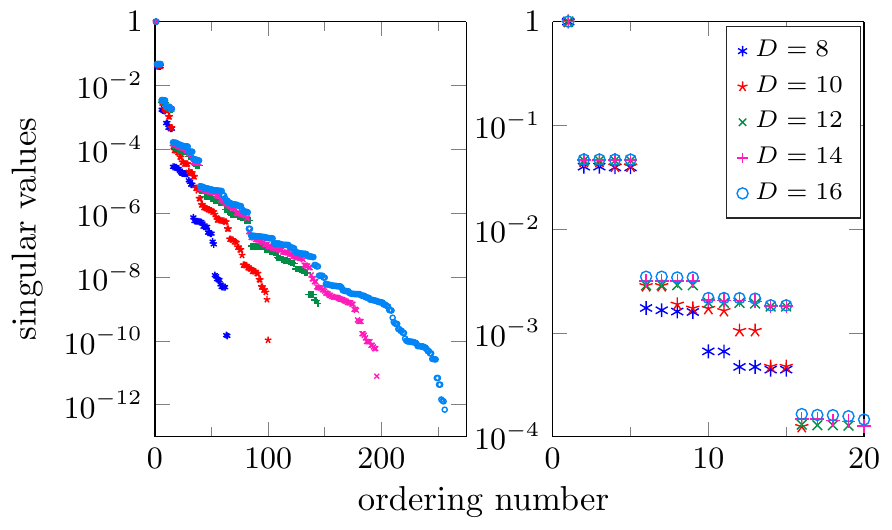}
\caption{
Spectrum of singular values of the renormalized $D'\times D'$ bond environment $e_l$ for a spinless half-filled Fermi sea ($\mu=0$ and $U=0$) at $\beta=2.56$. Here $D' = D^2$ and the largest singular value is normalized to 1.
The left panel shows all the -- quickly decaying -- singular value for each $D$ and the right panel focuses on the few largest ones demonstrating the convergence with increasing $D$.
}
\label{fig:eSVD}
\end{figure}

In this appendix, in order to provide better understanding and further justification of the introduced algorithm in Fig.~\ref{fig:eSVD} we show spectrum of singular values of the renormalized $D'\times D'$ bond environment $e_l$. It converges fast with growing $D'$ and decays rapidly demonstrating that a good approximation of the huge unrenormalized bond environment $E_l$ by the renormalized $e_l$ can be obtained  for large enough $D'$.

\FloatBarrier

%%%%%%%%%%%%%%%%%%%%%%%%%%%%%%%%%%%%%%%%%%%%%%%%%%%%%%%%%%%%%%%%%%%%%%%%%%%%

\end{document}